\documentclass[10pt,journal,twocolumn]{IEEEtran}
\IEEEoverridecommandlockouts
\usepackage{epsfig,cite,bm,algorithm,algorithmic,epstopdf,amsmath,subfigure,multirow,amssymb,amsfonts,amsthm,xcolor,caption,graphicx,textcomp,xcolor, subfigure, makecell, bbding}
\usepackage[inline, shortlabels]{enumitem}
\usepackage[flushleft]{threeparttable}
\setlist{itemjoin ={,\enspace},itemjoin* = { and\enspace}}

\usepackage[T1]{fontenc}
\usepackage{stfloats}
\allowdisplaybreaks[4]

\allowdisplaybreaks[4]
\usepackage{booktabs}

\begin{document}

\title{CRB-Rate Tradeoff in RSMA-enabled Near-Field Integrated Multi-Target Sensing and Multi-User Communications}
	
\author{Jiasi Zhou, Cong Zhou, Yanjing Sun,~\IEEEmembership{Member,~IEEE}, and Chintha Tellambura,~\IEEEmembership{Fellow,~IEEE}
\thanks{Jiasi Zhou is with the School of Medical Information and Engineering, Xuzhou Medical University, Xuzhou, 221004, China, (email: jiasi\_zhou@xzhmu.edu.cn). (\emph{Corresponding author: Jiasi Zhou}).}
\thanks{Cong Zhou is with the School of Electronic and Information Engineering, Harbin Institute of Technology, Harbin 150001, China, (email: zhoucong@stu.hit.edu.cn).}
\thanks{Yanjing Sun is with the School of Information and Control Engineering, China University of Mining and Technology, Xuzhou 221116, China (email: yjsun@cumt.edu.cn).}
\thanks{ Chintha Tellambura is with the Department of Electrical and Computer Engineering, University of Alberta, Edmonton, AB, T6G 2R3, Canada (email: ct4@ualberta.ca).} 
\thanks{This work was supported by the national key research and development program of China (2020YFC2006600) and the Talented Scientific Research Foundation of Xuzhou Medical University (D2022027).}}
\maketitle

\begin{abstract}
Extremely large-scale antenna arrays enhance spectral efficiency and spatial resolution in integrated sensing and communication (ISAC) networks while expanding the Rayleigh distance, triggering a shift from conventional far-field plane waves to near-field (NF) spherical waves. However, full-digital beamforming is infeasible due to the need for dedicated radio frequency (RF) chains. To address this, NF-ISAC with a rate-splitting multiple access (RSMA) scheme is developed for advanced interference management, considering fully-connected and partially-connected hybrid analog and digital (HAD) beamforming architectures. Specifically, the Cram\'{e}r-Rao bound (CRB)  for joint distance and angle sensing is derived, and the achievable performance region between the max-min communication rate and the multi-target CRB is defined. To fully characterize the Pareto boundary of the CRB-rate region, a sensing-centric minimization problem is formulated under communication rate constraints for two HAD beamforming architectures. A penalty dual decomposition (PDD)-based double-loop algorithm is developed to optimize fully-connected HAD beamformers. To reduce computational complexity, a two-stage design algorithm for fully connected HAD beamforming is also proposed. Additionally, the PDD-based double-loop algorithm is extended to the partially-connected HAD architecture. Simulations demonstrate the proposed schemes and algorithms: 1) achieve performance comparable to a fully digital beamformer with fewer RF chains, 2) outperform space division multiple access and far-field ISAC, and 3) yield enhanced CRB-rate trade-off performance.
\end{abstract} 

\begin{IEEEkeywords}
Near-field communications, integrated sensing and communications, rate splitting multiple access, Cram\'{e}r-Rao bound, multi-target sensing.
\end{IEEEkeywords}

\section{Introduction} 
Integrated sensing and communication (ISAC) unifies wireless communication and remote sensing through spectrum sharing and joint signal processing \cite{9737357}. To achieve ultra-high spectrum efficiency and spatial resolution, ISAC employs extremely large-scale antenna arrays (ELAA) operating at high frequencies, with potentially hundreds or thousands of antennas \cite{wang2024performance}. The increased antenna aperture and carrier frequency extend the Rayleigh distance to dozens or even hundreds of meters \cite{10587118}, shifting electromagnetic propagation from far-field (FF) plane waves to near-field (NF) spherical waves \cite{he2024unlocking}.

Unlike plane-wave channels, which only contain angular information and require significant bandwidth for accurate distance estimation\cite{10663521}, spherical-wave channels include an additional distance domain, enabling simultaneous angle and distance estimation with limited bandwidth\cite{10388218}. This capability opens new possibilities for multi-target sensing. However, conventional ISAC designs based on FF plane-wave models are unsuitable for NF sensing, necessitating new approaches tailored for NF-ISAC.

However, NF-ISAC may have complex interference issues due to the dual communication and sensing tasks \cite{10486996}. For example, in communication-only networks, interference signals are other user signals, \emph{i.e.}, inter-user interference. In contrast, the ISAC interference sources include communication beams from other users and additional sensing beams. Moreover, ISAC networks add sensing functions compared to communication networks.   When detecting multiple targets, the interference at the base station (BS) end stems from the transmit and echo signals due to the overlapping frequency bands. To limit interference, ISAC can adopt space division multiple access (SDMA)\cite{10050406} and non-orthogonal multiple access (NOMA)\cite{9927490,10423585}. SDMA leverages spatial multiplexing through beamforming to mitigate interference, but its effectiveness is limited as network data rates plateau when interference dominates\cite{mao2018rate}. NOMA employs successive interference cancellation (SIC) to decode and remove strong interference, enhancing throughput and connectivity. However, NOMA needs complex receiver designs, aligned user channels, and significant channel differences to achieve its performance gains \cite{mao2018rate}, limiting its applicability in ISAC.

To attack interference, one promising approach is rate-splitting multiple access (RSMA), which offers a flexible and robust interference management strategy \cite{10273395,10038476}. RSMA operates by splitting user messages into common and private parts, transmitting the common part to all users while decoding and canceling partial interference, and tolerating residual interference. By optimizing the common stream, RSMA effectively balances interference suppression and resource allocation. This approach generalizes both SDMA and NOMA, adapting dynamically to different interference levels and user deployments. Its scalability and efficiency make RSMA well-suited for various wireless scenarios. However, despite its potential, RSMA remains mainly unexplored in NF-ISAC networks.

\subsection{Related Works}
Prior relevant contributions can be divided into four categories, namely, SDMA-enabled FF-ISAC\cite{10464353,10679658,10251151,10382465}, RSMA-enabled FF-ISAC\cite{gong2024hybrid,10486996,10032141,10522473,10287099}, full digital beamforming for NF-ISAC\cite{10520715,10694020,hua2024near,10681603}, and suboptimal multiple access strategies for NF-ISAC\cite{10135096,meng2024hybrid,10700785,10579914}. It is evident that most works consider FF rather than NF propagation.  We next comprehensively survey the recent advances in these categories.

\subsubsection{SDMA-enabled FF-ISAC} SDMA-enabled FF-ISAC designs have been extensively studied \cite{10464353,10679658,10251151,10382465}. For instance, reference \cite{10464353} maximizes the minimum signal-to-interference-plus-noise ratio (SINR) for multi-target sensing, where a dual-function base station (BS) and intelligent omni surfaces (IOS) collaborate for 360-degree coverage. In \cite{10679658}, transmit beampattern and sensing SINR are optimized under imperfect channel state information (CSI). The authors in \cite{10251151} derive the Cram\'{e}r-Rao bound (CRB) of angle estimation and minimize the complete response matrix and reflection coefficients in multi-target ISAC over multicast channels, addressing target detection and tracking phases. This approach is further extended in \cite{10382465} to simultaneous wireless information and power transfer (SWIPT) systems, where the Pareto boundary of the CRB-rate-energy region is optimized via the transmit covariance matrix. Nonetheless, these studies primarily rely on transmit beamforming to mitigate interference, limiting network performance.

\subsubsection{RSMA-enabled FF-ISAC} To address performance saturation, the common stream of RSMA offers three key functions: mitigating interference between communication and sensing, managing interference among communication users, and acting as the sensing beam \cite{9531484}. This insight has spurred interest in RSMA-enabled FF-ISAC \cite{gong2024hybrid,10486996,10032141,10522473,10287099}. For instance, Reference \cite{gong2024hybrid} eliminates dedicated radar sequences in transmitted signals and designs HAD beamforming to maximize single-target sensing SINR. Extending this, reference \cite{10486996} adapts the transmit scheme to multi-target networks and optimizes CRB estimation, albeit using a full digital beamforming architecture. In \cite{10032141}, the tradeoff between communication and localization error is explored through cooperative multi-BS networks. RSMA is also integrated with reconfigurable intelligent surfaces (RIS) to enhance target detection SINR \cite{10287099} or reduce sensing error \cite{10522473} by jointly optimizing active and passive beamforming. However, these RSMA-enabled ISAC designs rely on FF plane-wave channels and predominantly use full digital beamforming, limiting joint distance and angle estimation capabilities.

\subsubsection{Full digital beamforming for NF-ISAC}
Although NF spherical waves can resolve distance and angle simultaneously, NF-ISAC remains underexplored, with notable exceptions in \cite{10520715,10694020,hua2024near,10681603,10135096,meng2024hybrid,10700785,10579914}. Existing studies design the transmit covariance matrix to optimize SINR \cite{10520715}, beampattern matching \cite{10694020}, and sum-CRB \cite{hua2024near} for multi-target detection while ensuring communication rate and power constraints. These three efforts utilize the semi-definite relaxation method to attack the formulated non-convex problems, incurring high computational complexity\cite{10520715,10694020,hua2024near}. Additionally, similar to FF-ISAC \cite{10287099,10522473}, NF-ISAC can deploy RIS to enhance capacity and coverage, but it uniquely enables joint distance and angle estimation \cite{10681603}. However, these NF-ISAC designs use full digital beamforming, which poses significant hardware challenges due to RF chain demands. 

 \subsubsection{Suboptimal multiple access strategies for NF-ISAC} To tradeoff the performance and hardware cost, the authors in \cite{10135096} introduce HAD beamforming and derive the sensing CRB for single-target ISAC. Similarly, SINR is optimized in \cite{meng2024hybrid} and \cite{10700785} 
 for single-target detection using HAD architectures. Additionally, a double-array transceiver structure is proposed in \cite{10579914}  for downlink and uplink ISAC. However, these contributions adopt suboptimal multiple access strategies, which cannot manage interference flexibly\cite{10135096, meng2024hybrid, 10700785, 10579914}. The RSMA-enabled NF-ISAC is underexplored, except for \cite{zhou2024hybrid}. Reference \cite{zhou2024hybrid} demonstrates that dedicated sensing beams are unnecessary for RSMA-enabled NF-ISAC. Based on this insight, the minimum communication rate is maximized by optimizing the receiver filter and hybrid beamformers while ensuring the sensing rate. To our knowledge, the tradeoff between CRB for joint distance and angle sensing and communication performance in RSMA-enabled NF-ISAC remains unexplored.

\subsection{Motivations and Contributions}
To address the research gaps, we propose an RSMA-enabled NF-ISAC framework for multi-target and multi-user scenarios, considering fully-connected and partially-connected HAD beamforming architectures. Table \ref{Table I} compares existing studies, highlighting the novelty of our work. The motivations are as follows:

\begin{itemize}
\item \textbf{Precise Interference Management}: As mentioned earlier,  current NF-ISAC designs use SDMA \cite{10520715,10694020,hua2024near,10681603,10135096,meng2024hybrid,10700785,10579914}, which cannot flexibly manage interference. In contrast, the RSMA common stream can mitigate communication-sensing interference, manage user interference, or serve as a sensing beam, enhancing network performance \cite{9531484}. This motivates the use of an RSMA-based transmit scheme for NF-ISAC.

\item \textbf{Reducing Hardware Complexity}: NF-ISAC networks rely on ELAAs and operate at high frequencies to extend Rayleigh distance, imposing significant RF chain deployment challenges \cite{10579914,10135096}. Fully digital beamforming becomes impractical as each antenna requires a dedicated RF chain, motivating the adoption of HAD beamforming architectures.

\item \textbf{Joint distance and angle estimation}: RSMA and NF-ISAC have been treated separately, except for our prior work \cite{zhou2024hybrid}, which uses the sensing rate as a metric. However, the resulting SINR optimization may not fully capture sensing quality, potentially compromising performance \cite{10050406}. In contrast, the CRB directly measures the sensing performance limits for target parameter estimation, providing a lower bound on the variance of unbiased estimators by leveraging the Fisher information matrix \cite{9705498}. Moreover, unlike FF plane waves, NF spherical waves include both distance and angle information, enabling CRB in NF-ISAC to be utilized for estimating both simultaneously. This motivates our adoption of the CRB as the sensing performance metric.
\end{itemize}

\begin{table*}[h]
\setlength{\tabcolsep}{5pt}
	\caption{Current ISAC studies vs this work.} 
	\begin{center}\label{Table I}
    \begin{threeparttable}
		\begin{tabular}{|c||c|c|c|c|c|c|c|c|c|c|c|c|c|c|} 
			\hline
&\cite{10464353,10679658}&\cite{10251151}&\cite{10382465}&\cite{gong2024hybrid}&\cite{10486996}& \cite{10032141,10522473}&\cite{10287099}&\cite{10520715,10694020}&\cite{hua2024near}&\cite{10681603}&\cite{10135096,10579914}&\cite{meng2024hybrid,10700785}&\cite{zhou2024hybrid}&\makecell*[c]{\bf{Our work}}\\
                \hline 
           	\makecell*[c]{\bf{NF/FF channel}} &FF&FF&FF&FF&FF&FF&FF&NF&NF&NF&NF&NF&NF&NF\\
                \hline 
           	\makecell*[c]{\bf{Multiple access}} &S  &S&S &R &R &R&R&S&S&S&S&S&R&R\\
                \hline 
           	\makecell*[c]{\bf{HAD beamforming }} && & &\Checkmark& &&&&&&\Checkmark&\Checkmark&\Checkmark&\Checkmark\\
	          \hline 
           	\makecell*[c]{\bf{Multiple targets}} & \Checkmark&\Checkmark &&&\Checkmark&& &\Checkmark&\Checkmark&&&&\Checkmark&\Checkmark\\
            \hline 
           	\makecell*[c]{\bf{CRB}} & &\Checkmark & \Checkmark& &\Checkmark&\Checkmark&&&\Checkmark&\Checkmark&\Checkmark&&&\Checkmark\\
                        \hline 
           	\makecell*[c]{\bf{Beampattern/SINR}} & \Checkmark& & &\Checkmark&&&\Checkmark&\Checkmark&&&&\Checkmark&\Checkmark&\\
	          \hline
		\end{tabular}
        \begin{tablenotes}
	\small {
	\item{ $\dagger$ - S and R denote SDMA and RSMA, respectively.}}
    \end{tablenotes}
\end{threeparttable}
	\end{center}
\end{table*}

Our main contributions are as follows:  

\begin{itemize}  
\item \textbf{RSMA-Based NF-ISAC Scheme}: We propose this by exploiting the HAD beamforming architectures. RSMA enables flexible interference management among users and dual functions, while fully-connected and partially-connected HAD beamforming reduces hardware complexity.  The multi-target CRB for joint distance and angle sensing is derived, and the CRB-rate achievable region is defined to capture the tradeoff between communication rate and sensing CRB. A dual-objective optimization problem is formulated.  

\item \textbf{Optimization Framework}: The dual-objective problem is simplified into a sensing-centric minimization problem with communication rate constraints. A penalty dual decomposition (PDD)-based dual-loop framework is developed to jointly optimize the fully-connected HAD beamformer and common rate allocation. The PDD framework employs an augmented Lagrangian (AL) formulation solved using block coordinate descent (BCD).  

\item \textbf{Low-Complexity Approach}: A two-stage algorithm is proposed for fully-connected HAD beamforming to avoid double-loop iterations, reducing computational complexity. Additionally, the PDD-based framework is adapted to handle partially-connected HAD beamforming.  

\item \textbf{Performance Insights}: Simulations demonstrate three key advantages: (1) performance comparable to full digital beamforming with fewer RF chains, (2) significant gains over SDMA and FF ISAC, and (3) enhanced CRB-rate tradeoff performance across benchmarks.  
\end{itemize}

\emph{Organization:}   
Section \ref{Section II} describes the system model and formulates the optimization problem. Section \ref{Section III} presents the PDD-based double-loop and low-complexity two-stage fully connected HAD beamforming algorithms, while Section \ref{Section IV} extends the proposed PDD-based double-loop algorithm to partially connected HAD beamforming. Section \ref{Section V} discusses simulation results, and Section \ref{Section VI} concludes the paper.

\emph{Notations:} Boldface upper-case letters, boldface lower-case letters, and calligraphy letters denote matrices, vectors, and sets. The $N\times K$ dimensional complex matrix space is denoted by $\mathbb{C}^{N\times K}$. The superscripts ${(\bullet)}^T$, ${(\bullet)}^*$, and ${(\bullet)}^H$ represent the transpose, conjugate, and Hermitian transpose, respectively. $\text{Re}\left( \bullet\right)$, $\text{Tr}\left( \bullet\right)$,  $\text{rank}\left( \bullet\right)$, and $\mathbb{E}\left[\bullet\right]$ 
 denote the real part, trace, rank, and statistical expectation. $\text{diag}\left( \bullet\right)$  and $\text{Bdiag}\left( \bullet\right)$ denote the diagonal and block diagonal operations, respectively. Operator $\lfloor a\rfloor$  is the largest integer not greater than $a$. $\odot$ and $\otimes$ denote the Hadamard and Kronecker products. Variable  $x\sim\mathcal{CN}(\mu, \sigma^2)$ is a  circularly symmetric complex Gaussian (CSCG)   with mean $\mu$ and variance $\sigma^2$.

\section{System Model and Problem Formulation}\label{Section II}
\begin{figure}[tbp]
\centering
\includegraphics[scale=0.9]{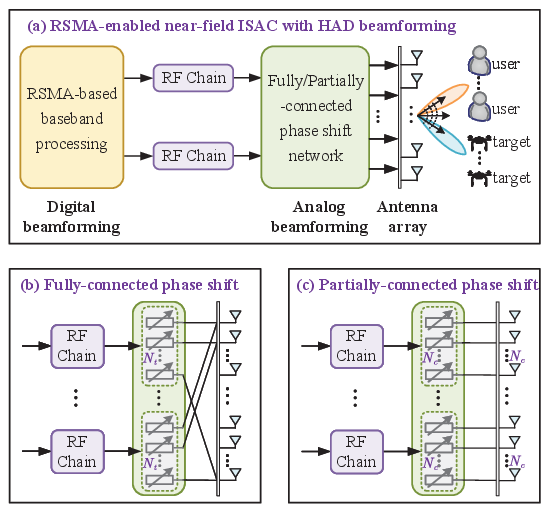}
\caption{The considered RSMA-aided NF-ISAC networks.}
\label{fig: system}
\end{figure}
As mentioned before, this study proposes an RSMA-based transmit scheme for NF-ISAC networks,  Fig.~\ref{fig: system}(a), where a base station (BS) has $N_{\text{t}}$-transmit and $N_{\text{r}}$-receive antennas. It simultaneously communicates with $K$ single-antenna users and senses $M$ targets. The dual-function BS employs uniform linear arrays (ULA) with an inter-antenna spacing of $d$, resulting in array apertures of $D_i=(N_i-1)d$ for $\forall i\in\{\text{t},\text{r}\}$.  Let $\mathcal{K}=\{1,\dots, K\}$ and $\mathcal{M}=\{1,\dots, M\}$ denote the communication users and sensing targets sets. All users and targets are assumed to be located in the NF region. This indicates that the distances between the BS and users/targets are shorter than the Rayleigh distance $\frac{2D^2_i}{\lambda}$, where $\lambda$ is the signal wavelength. To reduce hardware overloads, HAD beamforming architecture is adopted, where a phase-shifted analog beamformer is installed between $N_{\text{f}}$ ($N_{\text{f}}<N_{\text{t}}$) RF chains and $N_{\text{t}}$ transmit antennas. This paper considers two different RF network architectures, which are described as follows,
\begin{itemize}
\item Fully-connected phase shift: each RF chain is attached to all the transmit antennas,  Fig.~\ref{fig: system}(b). Each entry of the analog beamformer $\mathbf{F}\in\mathbb{C}^{N_{\text{t}}\times N_{\text{f}}}$ meets the unit-modulus constraint, \emph{i.e.}, 
\begin{equation}
\mathcal{F}_{\text{FC}}=\left\{\mathbf{F}\big|\left|\mathbf{F}_{n,i}\right|=1,~n\in\mathcal{N}_{\text{t}},~i\in\mathcal{N}_{\text{f}}\right\},
\end{equation}
where $\mathcal{N}_{\text{t}}=\{1,\dots,N_{\text{t}}\}$ and $\mathcal{N}_{\text{f}}=\{1,\dots,N_\text{f}\}$. 

\item Partially-connected phase shift: each RF chain is connected to a $N_{\text{c}}$-antenna sub-array with $N_{\text{c}}=\frac{N_{\text{t}}}{N_{\text{f}}}$ ($N_{\text{c}}$ is assumed an integer for brevity),  Fig.~\ref{fig: system}(c). The analog beamformer is given by
\begin{equation}
\mathcal{F}_{\text{PC}}=\left\{\mathbf{F}\big|\mathbf{F}=\text{Bdiag}\left(\mathbf{f}_1,\dots,\mathbf{f}_{N_{\text{f}}}\right)\in\mathbb{C}^{N_{\text{t}}\times N_{\text{f}}}\right\},
\end{equation}
where  $\mathbf{f}_{i}\in\mathbb{C}^{N_{\text{c}}\times 1}$ is a $N_{\text{c}}$-dimension vector with unit-modulus constraint, \emph{i.e.}, $\left|\mathbf{f}_{i,j}\right|=1$ for $i\in\mathcal{N}_{\text{f}}$ and $j\in\mathcal{N}_\text{c}=\{1,\dots,N_{\text{c}}\}$, where $\mathbf{f}_{i,j}$ denotes the $j$-th element of column vector $\mathbf{f}_i$.
\end{itemize}

\subsection{NF communication and sensing channel models}
Without loss of generality, the transmit and receive ULAs are placed along the positive and negative $y$-axis, respectively. Hence, the coordinate of the $n$-th transmit antenna is $\mathbf{s}_n=\left(0, nd\right)$, where $n\in\mathcal{N}_{\text{t}}$. Let $r_k$ and $\theta_k$ denote the distance and angle of the $k$-th user, so its coordinate is $\mathbf{r}_k=\left(r_k\cos\theta_k,r_k\sin\theta_k\right)$. The distance between the $n$-th transmit antenna and user $k$ is 
\begin{equation}
d_{k,n}=||\mathbf{r}_k-\mathbf{s}_n|| = \sqrt{r^2_k+( nd)^2-2ndr_k\sin\theta_k}.
\end{equation}
Furthermore, the  Fresnel approximation suggests that the path loss for the channels between all transmit antennae and the $k$-th user is identical\cite{10220205}. As such, the free space path loss can be written as $\overline\beta_k=\frac{c}{4\pi fr_k}$, where $f$ and $c$ are the carrier frequency and speed of light, respectively. Consequently, the line-of-sight (LoS) channel between the $n$-th transmit antenna and the user can be represented as $h_{k,n}= \overline\beta_k e^{-j\frac{2\pi}{\lambda}d_{k,n}}$. Then, by using the second-order Taylor expansion $\sqrt{1+x}\approx 1+\frac{1}{2}x-\frac{1}{8}x^2$,  $d_{k,n}$ can be approximated as $d_{k,n}\approx r_k-\delta_{k,n}$, where $\delta_{k,n}=nd\sin\theta_k-(nd)^2\cos^2\theta_k/2r_k$\cite{10517348}. As a result,  $h_{k,n}\approx \beta_k e^{j\frac{2\pi}{\lambda}\delta_{k,n}}$, where $\beta_k=\overline\beta_ke^{-j\frac{2\pi}{\lambda}r_k}$. The NF LoS channel $\mathbf{h}_k\in\mathbb{C}^{N_{\text{t}}\times 1}$ between the BS and the $k$-user can be modeled as
\begin{align}
\mathbf{h}_k= \beta_k\big[e^{j\frac{2\pi}{\lambda}\delta_{k,1}},\dots,e^{j\frac{2\pi}{\lambda}\delta_{k,N_{\text{t}}}}\big]^T=\beta_k\mathbf{a}\left(d_k,\theta_k\right),
\label{Response_vector}
\end{align}
where $\mathbf{a}\left(d_k,\theta_k\right)$ is the NF array response vector.

This paper considers a general multi-path channel model, comprising a LoS channel and $Q$ non-LoS (NLoS) channels induced by scatters. Similar to the LoS channel model, the NLoS channels can be characterized. Then, the overall channel between the BS and the $k$-th user can be modeled as
\begin{equation}
\mathbf{h}_k= \beta_k\mathbf{a}\left(r_k,\theta_k\right) + \sum_{q=1}^{Q}\beta_{k,q}\mathbf{a}\left(r_{k,q},\theta_{k,q}\right),
\label{Overall_Channel}
\end{equation}
with $\beta_{k,q}=\overline\beta_{k,q} e^{-j\frac{2\pi}{\lambda}\left(r_{k,q}+\overline r_{k,q}\right)}$, where $r_{k,q} \left(\overline r_{k,q}\right)$ is the distance between the BS (the $k$-th user) and the $q$-th scatter associated to user $k$. Additionally, $\theta_{k,q}$ denotes the angle of the $q$-th scatter associated to user $k$. 

The NF sensing channels are modeled next. Target sensing depends on the echo signal received at the BS, so the round-trip channel needs to be considered. Let $\tilde r_m$, $\tilde \theta_m$, and $\tilde\beta_m$  denote the distance, angle, and complex channel gain of the $m$-th sensing target. Similar to the modeling approach of the communication channel, the sensing channel matrix $\mathbf{G}_m\in\mathbb{C}^{N_{\text{r}}\times N_{\text{t}}}$ between the BS and the $m$-th target is given by
\begin{equation}
\mathbf{G}_m=\tilde\beta_m\mathbf{a}_\text{r}\left(\tilde r_m,\tilde\theta_m\right)\mathbf{a}_\text{t}^T\left(\tilde r_m,\tilde\theta_m\right),
\end{equation}
where $\mathbf{a}_\text{r}\big(\tilde r_m,\tilde\theta_m\big)\in\mathbb{C}^{N_{\text{r}}\times 1}$ and $\mathbf{a}_\text{t}\big(\tilde r_m,\tilde\theta_m\big)\in\mathbb{C}^{N_{\text{t}}\times 1}$ denote the receive and transmit NF array response vector of the $m$-th target, respectively.

\subsection{Communication  and sensing metrics}
In the proposed  RSMA-based transmit scheme, the message $W_k$ of user $k$ is split into a common part $W_{c,k}$ and a private part $W_{p,k}$ for $\forall k\in\mathcal{K}$. The $K$ common parts  $\left\{W_{c,1},\dots,W_{c,K}\right\}$ are combined into a  common message $W_c$, which is encoded into one common stream $s_{0}$. Meanwhile, the $K$ private parts $\left\{W_{p,1},\dots, W_{p,K}\right\}$ are independently encoded into $K$ private streams $\left\{s_{1},\dots,s_{K}\right\}$. Let $L$ denote the discrete-time index within one coherent processing interval (CPI). The transmitted stream vector at time index $l$ is $\mathbf{s}\left(l\right)=\left[s_0(l),s_1(l),\dots,s_K(l)\right]^T\in\mathbb{C}^{(K+1)\times 1}$ for $\forall l\in\mathcal{L}=\{1,\dots, L\}$. RSMA with and without additional sensing beams presents the same tradeoff performance, eliminating the necessity for dedicated sensing sequences\cite{9531484}. The streams are linearly precoded by HAD beamformer $\mathbf{F}\mathbf{W}\in\mathbb{C}^{N_{\text{t}}\times (K+1)}$, where $\mathbf{W}=\left[\mathbf{w}_0,\mathbf{w}_1,\dots,\mathbf{w}_K\right]\in\mathbb{C}^{N_{\text{f}}\times (K+1)}$ is digital beamforming matrix. $\mathbf{w}_0\in\mathbb{C}^{N_{\text{f}}\times 1}$ and $\mathbf{w}_k\in\mathbb{C}^{N_{\text{f}}\times 1}$ in $\mathbf{W}$ are designed for the common stream and $k$-th private stream, respectively. The transmit signal at $l$-th time index is  
\begin{equation}
\mathbf{x}\left(l\right)=\mathbf{F}\mathbf{W}\mathbf{s}(l)=\mathbf{F}\mathbf{w}_0s_0\left(l\right) + \sum_{k=1}^{K}{\mathbf{F}\mathbf{w}_ks_k\left(l\right)},
\label{transmit_signal}
\end{equation}
where the data streams meet $\frac{1}{L}\sum_{l=1}^{L}\mathbf{s}(l)\mathbf{s}^H(l)=\mathbf{I}_{K+1}$, indicating that the entries are mutual independent. Therefore, the covariance matrix of the transmit signal can be written as
\begin{equation}
\mathbf{R}=\frac{1}{L}\sum_{l=1}^{L}\mathbf{x}(l)\mathbf{x}^H(l) =\mathbf{F}\mathbf{W}\mathbf{W}^H\mathbf{F}^H.
\end{equation}

Based on equation (\ref{transmit_signal}), the received signal of the $k$-th user at  time index $l$ can be expressed as 
\begin{equation}
y_k\left(l\right)=\mathbf{h}^H_k\mathbf{F}\mathbf{w}_0s_0\left(l\right) + \mathbf{h}^H_k\sum_{k=1}^{K}{\mathbf{F}\mathbf{w}_ks_k\left(l\right)}+n_k,
\label{Receive_signal}
\end{equation}
where $n_k\sim \mathcal{CN}\left(0,\sigma^2_k\right)$ is the additional white Gaussian noise (AWGN) term. As such, the average power the $k$-th user receives is written as 
\begin{equation}
T_{c,k}=\overbrace{{\left|\mathbf{h}^H_{k}\mathbf{F}\mathbf{w}_0\right|}^2}^{S_{c,k}}+\underbrace{\overbrace{{\left|\mathbf{h}^H_{k}\mathbf{F}\mathbf{w}_{k}\right|}^2}^{S_{{p,k}}}+\overbrace{\sum_{j=1,j\neq k}^{K}\left|\mathbf{h}^H_{k}\mathbf{F}\mathbf{w}_j\right|^2+\sigma^2}^{I_{{p,k}}}}_{I_{c,k}=T_{p,k}}.
\label{equ:rece_power}
\end{equation}

To retrieve the intended message, user $k$ firstly decodes the common stream $s_0$ by treating $K$ private streams as noise, whose decoding SINR is $\gamma_{c,k}=S_{c,k}{I^{-1}_{c,k}}$. After that, the $k$-th user subtracts and removes the common message and decodes the private message $s_k$. The corresponding SINR is $\gamma_{{p,k}}=S_{p,k}{I^{-1}_{{p,k}}}$. As a result, the achievable rate of $s_0$ and $s_k$ at user $k$ are $R_{c,k}=\log\left(1 + \gamma_{{c,k}}\right)$ and $R_{p,k}=\log\left(1 + \gamma_{{p,k}}\right)$, respectively. However, to ensure that all $K$ users decode the common message successfully, the common rate cannot exceed $R_c= \min_{\forall k}R_{c,k}$. Meanwhile, since all users share the common rate, one has $R_c=\sum_{k=1}^{K}C_{c,k}$, where $C_{c,k}$ denotes the portion of the common rate allocated to user $k$. The total achievable rate of user $k$ is calculated as $R_k=C_{c,k}+R_{p,k}$.

In sensing space, the received echo signal from all targets at time block $l$ is 
\begin{equation}
\mathbf{y}\left(l\right) =\sum_{m=1}^{M}\mathbf{G}_m\mathbf{x}\left(l\right)+\mathbf{G}_{\text{SI}}\mathbf{x}\left(l\right)+\mathbf{n}_0,
\label{echo_signal}
\end{equation}
where $\mathbf{n}_0\sim \mathcal{CN}\left(0,\sigma^2_0\mathbf{I}_{N_{\text{r}}}\right)$ is AWGN and $\mathbf{G}_{\text{SI}}\in\mathbb{C}^{N_{\text{r}}\times N_{\text{t}}}$ is self-interference channel. Similar to \cite{10579914,10135096}, this paper assumes that self-interference can be perfectly canceled. As such, equation (\ref{echo_signal}) can be rewritten over $L$ time blocks as
\begin{equation}
\mathbf{Y}=\mathbf{A}_\text{r}\mathbf{B}\mathbf{A}^T_\text{t}\mathbf{X}+\mathbf{N}_0
\label{Sensing_matrix}
\end{equation}
where $\mathbf{Y}=\left[\mathbf{y}(1),\dots,\mathbf{y}(L)\right]$, $\mathbf{N}_0=\left[\mathbf{n}_0(1),\dots,\mathbf{n}_0(L)\right]$, $\mathbf{A}_\text{r}=\left[\mathbf{a}_\text{r}\big(\tilde r_1,\tilde\theta_1\big),\dots,\mathbf{a}_\text{r}\big(\tilde r_M,\tilde\theta_M\big)\right]$,  $\boldsymbol{\beta}=\left[\tilde\beta_1,\dots,\tilde\beta_M\right]$, $\mathbf{B}=\text{diag}\left(\boldsymbol{\beta}\right)$, and $\mathbf{A}_\text{t}=\left[\mathbf{a}_\text{t}\big(\tilde r_1,\tilde\theta_1\big),\dots,\mathbf{a}_\text{t}\big(\tilde r_M,\tilde\theta_M\big)\right]$.

This paper focuses on target tracking scenarios, where the BS is interested in estimating $\tilde \theta_m$, $\tilde r_m$, and $\tilde \beta_m$ based on the observation $\mathbf{Y}$ for $\forall m$. This can be achieved by the classic maximum likelihood estimation (MLE) algorithm. Readers are referred to \cite{10050406} for more details about the MLE algorithm. After that, the mean-squared error (MSE) is commonly used to evaluate the estimation performance, but optimizing MSE appears intractable. As a remedy, the CRB is adopted as the performance metric for target sensing, providing a lower MSE bound.  
Let $\boldsymbol{\beta}_\text{R}=\left[\text{Re}\big(\tilde\beta_1\big),\dots,\text{Re}\big(\tilde\beta_M\big)\right]^T$ and $\boldsymbol{\beta}_\text{I}=\left[\text{Im}\big(\tilde\beta_1\big),\dots,\text{Im}\big(\tilde\beta_M\big)\right]^T$ denote the real and imaginary parts of complex channel gain. Accordingly, there are   $4M$ unknown real parameters to be estimated, given by $\boldsymbol{\xi}=\left[\mathbf{r}^T,\boldsymbol{\theta}^T,\boldsymbol{\beta}^T_\text{R},\boldsymbol{\beta}^T_\text{I}\right]^T$, where $\mathbf{r}=\left[\tilde r_1,\dots,\tilde r_M\right]^T$ and $\boldsymbol{\theta}=\left[\tilde \theta_1,\dots,\tilde \theta_M\right]^T$. Then, the Fisher information matrix (FIM) for estimating vector $\bm{\xi}$ can be partitioned as
\begin{equation}
\mathbf{J}_{\bm{\xi}}= \frac{2L}{\sigma^2_0}\begin{bmatrix} \mathbf{J}_{11} & \mathbf{J}_{12} \\ \mathbf{J}^T_{12} & \mathbf{J}_{22} \end{bmatrix}\in\mathbb{R}^{4M\times 4M},
\label{FIM}
\end{equation}
where the expressions of matrices $\mathbf{J}_{11}\in\mathbb{R}^{2M\times 2M}$, $\mathbf{J}_{12}\in\mathbb{R}^{2M\times 2M}$, and $\mathbf{J}_{22}\in\mathbb{R}^{2M\times 2M}$ are derived in Appendix A. According to equation (\ref{FIM}), the CRB matrix for estimating $\mathbf{r}$ and $\bm{\theta}$ can be expressed as 
\begin{equation}
\text{CRB}\left(\mathbf{r},\bm{\theta}\right) = \frac{\sigma^2_0}{2L}\left(\mathbf{J}_{11}-\mathbf{J}_{12}\mathbf{J}^{-1}_{22}\mathbf{J}^T_{12}\right)^{-1}\in\mathbb{R}^{2M\times 2M}.\label{CRB}
\end{equation}

\subsection{Problem formulation}
The aim is to achieve  an optimal tradeoff between communication and sensing performance by jointly optimizing HAD beamformers and common rate allocation. This involves characterizing the CRB-rate achievable region by determining its Pareto boundary\footnote{The Pareto boundary represents the set of points where improving one performance metric requires compromising the other.}. Toward this end, this paper uses 
$\min_{\forall k}R_k$ and $1/\text{Tr}\left(\text{CRB}\left(\mathbf{r},\bm{\theta}\right)\right)$ as performance metrics to construct CRB-rate achievable region due to two reasons. First, $\min_{\forall k}R_k$ avoids prioritizing the transmit rate in a few links with favorable channel conditions, ensuring user fairness. Second, the reciprocal of CRB provides a closed region and lies in the first quadrant of the CRB-rate plane. Therefore, this problem is formulated as
\begin{subequations}\label{linear_p}
	\begin{align}
&\max_{\mathbf{F},\mathbf{W},\mathbf{c} } \left\{\left(\hat{R},\hat{\Phi}\right) \Big|\hat{R}\leq\min_{\forall k}R_k, \hat{\Phi}\leq  1/\text{Tr}\left(\text{CRB}\left(\mathbf{r},\bm{\theta}\right)\right)\right\},\label{ob_a}\\
	\text{s.t.}~
	&||\mathbf{F}\mathbf{W}||^2\leq P_{\text{th}},\label{ob_b}\\
 &\sum_{k=1}^{K}C_{c,k} \leq R_c,\label{ob_c}\\
 &C_{c,k} \geq 0, \quad k\in\mathcal{K},\label{ob_d}\\
  &|\mathbf{F}_{n,i}|\in\mathcal{F}_{x},~ n\in\mathcal{N}_{\text{t}},i\in\mathcal{N}_{\text{f}}, x\in\left\{\text{FC},\text{PC}\right\},\label{ob_e} 
	\end{align}
\end{subequations}
where $\mathbf{c}=\left[C_{c,1},\dots, C_{c,K}\right]^T$ and $P_{\text{th}}$ is the maximum transmit power threshold. (\ref{ob_b}) is the transmit power constraint.  (\ref{ob_c}) and (\ref{ob_d}) are the common rate allocation constraints. (\ref{ob_e}) is the analog beamformer constraint, where $x\in\left\{\text{FC},\text{PC}\right\}$ indicates different phase shift architectures in RF chains. 

Solving problem (\ref{linear_p}) is challenging for three technical reasons. First, problem (\ref{linear_p}) has dual-objective. Tradeoff performance analysis is difficult because of the complicated coupling between dual functionalities. Second, communication rate $R_k$ and sensing performance $\text{Tr}\left(\text{CRB}\left(\mathbf{r},\bm{\theta}\right)\right)$ are non-convex, making it hard to solve in the primary domain. Meanwhile, solving such problems is difficult in the dual domain due to the unknown duality gap. Third, analog and digital beamformers are highly coupled, aggravating the solution difficulty. Consequently, the global optimal solution appears elusive.

\section{Beamforming design for fully-connected phase shift architecture}\label{Section III}
This section focuses on HAD beamforming design for fully-connected phase shift architecture. Specifically, the dual-objective problem (\ref{linear_p}) is reformulated to a sensing-centric single-objective optimization problem. To attack the recast problem, a PDD-based double-loop algorithm is developed, in which the weighted minimum mean-squared error (WMMSE) approach is employed to recast the communication rate into easily optimized constraints. Then, a two-stage HAD beamforming strategy is proposed to reduce the computational complexity further. Finally,  critical properties of the proposed algorithms, such as convergence and complexity, are discussed. 

Two standard approaches can be adopted to depict the Pareto boundary: the weighting method and the constrained method. Many ISAC networks optimize transmit schemes by maximizing the weighted sum of dual functional metrics\cite{loli2022rate,9531484}. However, since the objective function and feasible set in problem (\ref{linear_p}) are non-convex, the weighting method cannot search for the complete Pareto boundary\cite{6565404}. This motivates us to utilize another approach to characterize the whole Pareto boundary. Specifically, we transform problem (\ref{linear_p}) into a sensing-centric problem, where the sensing error is minimized with constrained communication performance. The new problem is formulated as
\begin{subequations}\label{linear_p2}
	\begin{align}
&\min_{\mathbf{F},\mathbf{W},\mathbf{c} } \text{Tr}\left(\text{CRB}\left(\mathbf{r},\bm{\theta}\right)\right),\label{ob_a2}\\
	\text{s.t.}~
	&R_k\geq R_{\text{th}},~k\in\mathcal{K},\label{ob_b2}\\
   &|\mathbf{F}_{n,i}|=1,~ n\in\mathcal{N}_{\text{t}},i\in\mathcal{N}_{\text{f}},\label{ob_c2}\\
 &\mbox{(\ref{ob_b}) -- (\ref{ob_d})}.\label{ob_d2} 
	\end{align}
\end{subequations}
Solving problem (\ref{linear_p2}) for a given rate threshold $R_{\text{th}}$ yields a single point on the Pareto boundary of the CRB-rate region. Exhausting all possible values of $R_{\text{th}}$  characterizes the entire Pareto boundary. Therefore, the communication-centric optimization problem determines the complete Pareto boundary. 

\emph{Remark 1}: Before proceeding to solve the problem (\ref{linear_p2}), it is worth mentioning a special case with sole multi-user communication, which determines the maximum of $R_{\text{th}}$. The max-min fairness rate problem can be reduced to
\begin{subequations}\label{linear_p3}
	\begin{align}
&\max_{\mathbf{F},\mathbf{W},\mathbf{c},R_{\text{th}} } R_{\text{th}},\label{ob_a2}\\
	\text{s.t.}~
	 &\mbox{ (\ref{ob_b}) -- (\ref{ob_e}),~ (\ref{ob_b2})}.\label{ob_b3} 
	\end{align}
\end{subequations}
Problem (\ref{linear_p2}) degrades to problem (\ref{linear_p3}) when sensing performance is ignored. Therefore, it can be easily resolved if the problem (\ref{linear_p2}) is attacked.

\subsection{PDD framework for solving problem (\ref{linear_p2}) }
To eliminate the coupling between $\mathbf{F}$ and $\mathbf{W}$ in sensing error $\text{Tr}\left(\text{CRB}\left(\mathbf{r},\bm{\theta}\right)\right)$ and transmit rate $R_k$, we define two auxiliary matrice $\mathbf{P}=\mathbf{FW}$ and $\mathbf{Q}=\mathbf{P}\mathbf{P}^H$, where $\mathbf{P}=\left[\mathbf{p}_0,\dots,\mathbf{p}_k\right]\in\mathbb{C}^{N_{\text{t}}\times (K+1)}$. Substituting $\mathbf{P}=\mathbf{FW}$ into equation (\ref{equ:rece_power}), the average power can be written as
\begin{equation}
T_{c,k}=\overbrace{{\left|\mathbf{h}^H_{k}\mathbf{p}_0\right|}^2}^{S_{c,k}}+\underbrace{\overbrace{{\left|\mathbf{h}^H_{k}\mathbf{p}_{k}\right|}^2}^{S_{{p,k}}}+\overbrace{\sum_{j=1,j\neq k}^{K}\left|\mathbf{h}^H_{k}\mathbf{p}_j\right|^2+\sigma^2}^{I_{{p,k}}}}_{I_{c,k}=T_{p,k}},
\label{equ:rece_power2}
\end{equation}
Using $\mathbf{Q}=\mathbf{PP}^H$, the FIM in equation (\ref{FIM}) is updated to 
\begin{subequations}\label{partial_derivative2}
\begin{align}
\mathbf{J}_{11}=& \text{Re}\left(\begin{bmatrix} \mathbf{g}_{\bm{\theta}\bm{\theta}} & \mathbf{g}_{\bm{\theta}\bm{r}} \\ \mathbf{g}_{\bm{\theta}\bm{r}} & {\mathbf{g}_{\bm{r}\bm{r}}} \end{bmatrix}\right),\\
\mathbf{J}_{12}=&\text{Re}\left(\begin{bmatrix} \mathbf{g}_{\bm{\theta}\bm{\beta}_\text{R}}  \\ \mathbf{g}_{\bm{r}\bm{\beta}_\text{R}}  \end{bmatrix}\begin{bmatrix}1,j\end{bmatrix}\right),\\
\mathbf{J}_{22}=&\text{Re}\left(\begin{bmatrix} \mathbf{g}_{\bm{\beta}_\text{R}\bm{\beta}_\text{R}}& j\mathbf{g}_{\bm{\beta}_\text{R}\bm{\beta}_\text{R}}\ \\ j\mathbf{g}_{\bm{\beta}_\text{R}\bm{\beta}_\text{R}}\ & \mathbf{g}_{\bm{\beta}_\text{R}\bm{\beta}_\text{R}}\ \end{bmatrix}\right),
\end{align}
\end{subequations}
where
\begin{subequations}\label{partial_derivative3}
\begin{align}
\mathbf{g}_{\bm{x}\bm{y}}=&\left(\dot{\mathbf{A}}^H_{\text{r},\bm{x}}\dot{\mathbf{A}}_{\text{r},\bm{y}}\right)\odot\left(\mathbf{B}^*\mathbf{A}^H_\text{t}\mathbf{Q}^*\mathbf{A}_\text{t}\mathbf{B}\right)\notag\\
&+\left(\dot{\mathbf{A}}^H_{\text{r},\bm{x}}\mathbf{A}_{\text{r}}\right)\odot\left(\mathbf{B}^*\mathbf{A}^H_t\mathbf{Q}^*\dot{\mathbf{A}}_{\text{t},\bm{y}}\mathbf{B}\right)\notag\\
&+\left(\mathbf{A}^H_{\text{r}}\dot{\mathbf{A}}_{\text{r},\bm{y}}\right)\odot\left(\mathbf{B}^*\dot{\mathbf{A}}^H_{\text{t},\bm{x}}\mathbf{Q}^*\mathbf{A}_\text{t}\mathbf{B}\right)\notag\\
&+\left(\mathbf{A}^H_{\text{r}}\mathbf{A}_{\text{r}}\right)\odot\left(\mathbf{B}^*\dot{\mathbf{A}}^H_{\text{t},\bm{x}}\mathbf{Q}^*\dot{\mathbf{A}}_{\text{t},\bm{y}}\mathbf{B}\right),\\
\mathbf{g}_{\bm{x}\bm{\beta}_\text{R}}=&\left(\dot{\mathbf{A}}^H_{\text{r},\bm{x}}\mathbf{A}_{\text{r}}\right)\odot\left(\mathbf{B}^*\mathbf{A}^H_\text{t}\mathbf{Q}^*\mathbf{A}_{\text{t}}\right)\notag\\
&+\left(\mathbf{A}^H_{\text{r}}\mathbf{A}_{\text{r}}\right)\odot\left(\mathbf{B}^*\dot{\mathbf{A}}^H_{\text{t},\bm{x}}\mathbf{Q}^*\mathbf{A}_{\text{t}}\right),\\
\mathbf{g}_{\bm{\beta}_\text{R}\bm{\beta}_\text{R}}=&\left(\mathbf{A}^H_{\text{r}}\mathbf{A}_{\text{r}}\right)\odot\left(\mathbf{A}^H_{\text{t}}\mathbf{Q}^*\mathbf{A}_{\text{t}}\right),
\end{align}
\end{subequations}
for any $ \mathbf{x},\mathbf{y}\in\{\bm{\theta},\bm{r}\}$.

Based on the Schur complement condition, the quadratic equality constraint $\mathbf{Q}=\mathbf{P}\mathbf{P}^H$ is equivalent to  
\begin{subequations}\label{equation_constraint}
\begin{align}
&\begin{bmatrix} \mathbf{Q} & \mathbf{P} \\ \mathbf{P}^H & \mathbf{I} \end{bmatrix}\succeq\mathbf{0},\label{equation_1}\\
 &\text{Tr}\left(\mathbf{Q}-\mathbf{P}\mathbf{P}^H\right)\leq 0,\label{equation_2}
\end{align}
\end{subequations}

In addition, the complex form of the CRB matrix hinders the optimization process. To attack this issue, we transform the objective function (\ref{linear_p2}) into an equivalent but more tractable form. Specifically, a positive definite auxiliary matrix $\mathbf{U}\in\mathbb{C}^{2M\times 2M}$ is introduced, which meets
\begin{equation}
\mathbf{J}_{11}-\mathbf{J}_{12}\mathbf{J}^{-1}_{22}\mathbf{J}^T_{12}\succeq \mathbf{U}\succeq\mathbf{0}.\label{CRB_2}
\end{equation}
Inequality (\ref{CRB_2}) can be divided into the following two constraints by using the Schur complement
\begin{subequations}\label{Schur_complement}
\begin{align}
&\begin{bmatrix} \mathbf{J}_{11}-\mathbf{U} & \mathbf{J}_{12} \\ \mathbf{J}^T_{12} & \mathbf{J}_{22} \end{bmatrix}\succeq\mathbf{0},\label{Schur_1}\\
 & \mathbf{U}\succeq\mathbf{0},\label{Schur_2}
\end{align}
\end{subequations}
Since $\text{Tr}\left(\mathbf{X}^{-1}\right)$ is monotonicity decreasing on the positive semidefinite matrix $\mathbf{X}$, $\min \text{Tr}\left(\text{CRB}\left(\mathbf{r},\bm{\theta}\right)\right)$ is equivalent to $\min \text{Tr}\left(\mathbf{U}^{-1}\right)$ given that inequality (\ref{Schur_complement}) is satisfied. Consequently, the resultant new problem is
\begin{subequations}\label{linear_p4}
	\begin{align}
&\min_{\mathbf{P},\mathbf{Q},\mathbf{F},\mathbf{W},\mathbf{U},\mathbf{c} } \text{Tr}\left(\mathbf{U}^{-1}\right),\label{ob_a4}\\
	\text{s.t.}~
	&\mathbf{P}=\mathbf{F}\mathbf{W},\label{ob_b4}\\
 &||\mathbf{P}||^2\leq P_{\text{th}},\label{ob_c4}\\
  &\mbox{ (\ref{ob_c}),~(\ref{ob_d}),~(\ref{ob_b2}),~(\ref{ob_c2}),~(\ref{equation_constraint}),~(\ref{Schur_complement})}.\label{ob_d4}
	\end{align}
\end{subequations}
Note that the sensing performance and transmit rate in problem (\ref{linear_p4}) should be calculated by equations (\ref{equ:rece_power2}) and (\ref{partial_derivative2}), respectively. To counter the equality constraint (\ref{ob_b4}), the PDD optimization framework is utilized in a double-loop structure. Specifically, problem (\ref{linear_p4}) is reconstructed as a new problem without equation constraint by introducing the Lagrangian dual matrix and penalty factor. Then, the augmented Lagrangian (AL) problem is optimized in the inner loop, while the introduced auxiliary variables are updated in the outer loop. By introducing the Lagrangian dual matrix $\mathbf{D}$ and penalty factor $\rho$, we can get the corresponding AL problem
\begin{subequations}\label{linear_p5}
	\begin{align}
&\min_{\mathbf{P},\mathbf{Q},\mathbf{F},\mathbf{W},\mathbf{U},\mathbf{c} } \text{Tr}\left(\mathbf{U}^{-1}\right)+\frac{1}{2\rho}||\mathbf{P}-\mathbf{FW} +\rho\mathbf{D}||^2,\label{ob_a5}\\
	\text{s.t.}~
  &\mbox{ (\ref{ob_c4}),~(\ref{ob_d4})}.\label{ob_d5}
	\end{align}
\end{subequations}
The proposed PDD-based double-loop algorithm for solving
problem (\ref{linear_p5}) is summarized in Alg.~\ref{Alg.1}. The detailed analysis and discussion on convergence and optimality of the PDD framework can be found in \cite{9120361}. We can easily validate that $\frac{1}{2\rho}||\mathbf{P}-\mathbf{FW} +\rho\mathbf{D}||^2\to 0$ when $\rho\to 0$, which indicates that equality constraint (\ref{ob_b4}) is satisfied. The major challenge of the proposed PDD-based algorithm lies in solving the AL problem (\ref{linear_p5}) in line 3. In the next subsection, we will illustrate the procedure of solving the AL problem (\ref{linear_p5}).
\begin{algorithm}[t]
	\caption{PDD-based double-loop algorithm for solving problem (\ref{linear_p5})}
	\begin{algorithmic}[1]\label{Alg.1}
		\STATE Initialize $\mathbf{F}^{(0)}$, $\mathbf{W}^{(0)}$, $\mathbf{D}^{(0)}$, $\rho^{(0)}$, and $\psi^{(0)}$, set iteration index $n=1$ and the maximum tolerance $\xi_1$.
		\WHILE{ not convergent}
		\STATE  Solving problem (\ref{linear_p5}) to obtain $\mathbf{P}^{(n)}$, $\mathbf{F}^{(n)}$, and $\mathbf{W}^{(n)}$.
        \IF{ $||\mathbf{P}^{(n)}-\mathbf{F}^{(n)}\mathbf{W}^{(n)}||_\infty\leq \psi^{(n-1)}$ }
		\STATE Update $\mathbf{D}^{(n)}=\mathbf{D}^{(n-1)}+\frac{1}{\rho^{(n)}}\left(\mathbf{P}^{(n)}-\mathbf{F}^{(n)}\mathbf{W}^{(n)}\right)$.
        \STATE Keep penalty factor unchanged, \emph{i.e.},~$\rho^{(n)}=\rho^{(n-1)}$.
        \ELSE
        \STATE Update $\rho^{(n)}=\mu\rho^{(n-1)}$, where $0<\mu<1$.
        \STATE Keep Lagrangian dual matrix $\mathbf{D}$ unchanged, \emph{i.e.}, $\mathbf{D}^{(n)}=\mathbf{D}^{(n-1)}$.
		\ENDIF
        \STATE Update $\psi^{(n)}=0.9||\mathbf{P}^{(n)}-\mathbf{F}^{(n)}\mathbf{W}^{(n)}||_\infty$ and $n=n+1$.
		\ENDWHILE	
		\STATE Output the optimal sensing performance.
	\end{algorithmic}
\end{algorithm}

\subsection{ BCD algorithm for solving AL problem (\ref{linear_p5})}
To separate coupling variables, the optimization variables in problem (\ref{linear_p5}) are divided into three blocks, namely, $\left\{\mathbf{P},\mathbf{Q},\mathbf{U},\mathbf{c}\right\}$, $\left\{\mathbf{W}\right\}$, and $\left\{\mathbf{F}\right\}$. Then, the BCD method is utilized to optimize one block alternately while keeping other blocks at their previous value\cite{10587118}. The optimization of each block is elaborated as follows.

\emph{1) Subproblem over $\left\{\mathbf{P},\mathbf{Q},\mathbf{U},\mathbf{c}\right\}$}: With fixed $\left\{\mathbf{F},\mathbf{W}\right\}$, the optimization problem for $\left\{\mathbf{P},\mathbf{Q},\mathbf{U},\mathbf{c}\right\}$ can be simplified as
\begin{subequations}\label{linear_p6}
	\begin{align}
&\min_{\mathbf{P},\mathbf{Q},\mathbf{U},\mathbf{c} } \text{Tr}\left(\mathbf{U}^{-1}\right)+\frac{1}{2\rho}||\mathbf{P}-\mathbf{FW} +\rho\mathbf{D}||^2,\label{ob_a6}\\
	\text{s.t.}~
  &\mbox{ (\ref{ob_c}),~(\ref{ob_d}),~(\ref{ob_b2}),~(\ref{equation_constraint}),~(\ref{Schur_complement})}.\label{ob_b6}
	\end{align}
\end{subequations} 
Problem (\ref{linear_p6}) has high non-convexity due to the fractional SINR and is thus generally difficult to solve optimally. To tackle this problem, we employ the WMMSE approach to reformulate the communication rate. Specifically, user $k$ employs equalizers $\omega_{c,k}$ and $\omega_{p,k}$ to detect the common stream $s_0$ and private stream $s_k$, respectively. Meanwhile, since the common stream has been removed before decoding the private stream, we have $\hat{s}_{c,k}=\omega_{c,k}y_k$ and $\hat{s}_{k}=\omega_{p,k}\left(y_k-\mathbf{h}^H_k\mathbf{p}_0s_0\right)$, where $\hat s_{c,k}$ and $\hat{s}_{k}$ are the estimation of $s_0$ and $s_k$, respectively. The MSEs of detecting $s_0$ and $s_k$ are respectively calculated as 
\begin{subequations}\label{MMSE_error}
\begin{align}
\delta_{c,k} = &\mathbb{E}\left\{\left|\hat{s}_{c,k}-s_{0}\right|^2\right\}\notag\\=&\left|\omega_{c,k}\right|^2T_{c,k}-2\text{Re}\left(\omega_{c,k}\mathbf{h}^H_k\mathbf{p}_0\right)+1,\\
\delta_{p,k} = &\mathbb{E}\left\{\left|\hat{s}_{k}-s_{k}\right|^2\right\}\notag\\=&\left|\omega_{p,k}\right|^2T_{p,k}-2\text{Re}\left(\omega_{p,k}\mathbf{h}^H_k\mathbf{p}_k\right)+1.
\end{align}
\end{subequations}
The optimum minimum MSE (MMSE) equalizers are then calculated by solving $\frac{\partial \delta_{c,k}}{\partial \omega_{c,k}}=0$ and $\frac{\partial \delta_{p,k}}{\partial \omega_{p,k}}=0$, which are given by 
\begin{align}
\omega^{\text{MMSE}}_{c,k}=\mathbf{p}^H_0\mathbf{h}_kT^{-1}_{c,k}, \quad \text{and }\quad\omega^{\text{MMSE}}_{p,k}=\mathbf{p}^H_k\mathbf{h}_kT^{-1}_{p,k}.
\label{Optimal_equalizer}
\end{align}
Substituting (\ref{Optimal_equalizer}) into (\ref{MMSE_error}), the MMSEs are 
\begin{subequations}
\begin{align}
\delta^{\text{MMSE}}_{c,k} =\min_{\omega_{c,k}}\delta_{c,k}= &T^{-1}_{c,k}I_{c,k},\\
\delta^{\text{MMSE}}_{p,k} =\min_{\omega_{p,k}}\delta_{p,k}= &T^{-1}_{p,k}I_{p,k}.
\label{MMSE_error2}
\end{align}
\end{subequations}
The SINRs of decoding $s_0$ and $s_k$ at user $k$ can be respectively transformed  to $\gamma_{c,k}=1/\delta^{\text{MMSE}}_{c,k}-1$ and $\gamma_{p,k}=1/\delta^{\text{MMSE}}_{p,k}-1$.  The corresponding transmit rates become $R_{c,k}=-\log\big(\delta^{\text{MMSE}}_{c,k}\big)$ and $R_{p,k}=-\log\big(\delta^{\text{MMSE}}_{p,k}\big)$.

By introducing the positive weights $\eta_{c,k}$ and $\eta_{p,k}$, the weighted MSEs (WMSEs) of decoding $s_0$ and $s_k$ at the $k$-th user are defined as 
\begin{subequations}\label{WMSE}
\begin{align}
\beta_{c,k}=&\eta_{c,k}\delta_{c,k}-\log\left(\eta_{c,k}\right), \\
\beta_{p,k}=&\eta_{p,k}\delta_{p,k}-\log\left(\eta_{p,k}\right).
\end{align}
\end{subequations}
The optimum equalizers for minimizing $\beta_{c,k}$ and $ \beta_{p,k}$ are respectively $\omega^o_{c,k}=\omega^{\text{MMSE}}_{c,k}$ and $\omega^o_{p,k}=\omega^{\text{MMSE}}_{p,k}$, which can be derived by solving $\frac{\partial \beta_{c,k}}{\partial \omega_{c,k}}=0$ and $\frac{\partial \beta_{p,k}}{\partial \omega_{p,k}}=0$. Substituting $\omega^{\text{MMSE}}_{c,k}$ and $\omega^{\text{MMSE}}_{p,k}$ into equation (\ref{WMSE}), we have
\begin{subequations}
\begin{align}
\beta_{c,k}=&\eta_{c,k}\delta^{\text{MMSE}}_{c,k}-\log\left(\eta_{c,k}\right), \\
\beta_{p,k}=&\eta_{p,k}\delta^{\text{MMSE}}_{p,k}-\log\left(\eta_{p,k}\right).
\end{align}
\end{subequations}
By solving $\frac{\partial \beta_{c,k}}{\partial \eta_{c,k}}=0$ and $\frac{\partial \beta_{p,k}}{\partial \eta_{p,k}}=0$, we have
\begin{align}
\eta^o_{c,k}=\left(\delta^{\text{MMSE}}_{c,k}\ln 2\right)^{-1}~\text{and} ~\eta^o_{p,k}=\left(\delta^{\text{MMSE}}_{p,k}\ln 2\right)^{-1}.
\label{Optimal_weights}
\end{align}
The rate-WMMSE relationships are established by substituting (\ref{Optimal_weights}) into WMSE, which are given by 
\begin{subequations}\label{Relationship}
\begin{align}
\beta^{\text{MMSE}}_{c,k} &=\min_{\eta_{c,k},\omega_{c,k}}\beta_{c,k}= \tau-R_{c,k},\\
\beta^{\text{MMSE}}_{p,k} &=\min_{\eta_{p,k},\omega_{p,k}}\beta_{p,k}= \tau-R_{p,k},
\end{align}
\end{subequations}
where $\tau=1/\ln 2 +\log(\ln 2)$.

By taking the first-order Taylor expansion to  $\text{Tr}\left(\mathbf{P}\mathbf{P}^H\right)$, the  inequality (\ref{equation_2}) can be recast as
\begin{equation}
2\text{Re}\left(\text{Tr}\left(\mathbf{P}^{(t)}\mathbf{P}^H\right)\right)-\text{Tr}\left(\mathbf{P}^{(t)}\left(\mathbf{P}^{(t)}\right)^H\right)\geq \text{Tr}\left(\mathbf{Q}\right),
\label{Surrogate}
\end{equation}
where $(t)$ is the last iteration's updated value. Based on the rate-WMMSE relationships (\ref{Relationship}) and inequality (\ref{Surrogate}), problem (\ref{linear_p6}) can be transformed into the WMMSE problem 
\begin{subequations}\label{linear_p7}
	\begin{align}
&\min_{\mathcal{Q}_1,\mathcal{Q}_2} \text{Tr}\left(\mathbf{U}^{-1}\right)+\frac{1}{2\rho}||\mathbf{P}-\mathbf{FW} +\rho\mathbf{D}||^2,\label{ob_a7}\\
	\text{s.t.}~
 &\sum_{k=1}^{K}C_{c,k} + \min_{\eta_{c,k},\omega_{c,k}}\beta_{c,k} \leq \tau,~\forall k,\label{ob_b7}\\
 &C_{c,k} - \min_{\eta_{p,k},\omega_{p,k}}\beta_{p,k} \geq R_{\text{th}}-\tau,~\forall k,\label{ob_c7}\\
 &\mbox{ (\ref{ob_d}),~(\ref{ob_b2}),~(\ref{ob_c2}),~(\ref{equation_1}),~(\ref{Schur_complement}),~(\ref{Surrogate})}.\label{ob_d7}
	\end{align}
\end{subequations}
where $\mathcal {Q}_1$ and $\mathcal {Q}_2$ collect respectively the inherent variables in problem (\ref{linear_p6}) and introduced auxiliary variables, \emph{i.e.}, $\mathcal {Q}_1=\big\{\mathbf{P},\mathbf{Q},\mathbf{U},\mathbf{c}\big\}$ and $\mathcal {Q}_2=\big\{\eta_{c,k}, \omega_{c,k}, \eta_{p,k}, \omega_{p,k}\big\}$. For given auxiliary variables $\mathcal{Q}_2$, problem (\ref{linear_p7}) becomes convex w.r.t $\mathcal{Q}_1$, which can be effectively solved by standard convex optimization solver such as CVX\cite{cvx}. Driven by this observation, $\mathcal {Q}_1$ and $\mathcal {Q}_2$ are iteratively updated, where $\mathcal {Q}_2$ is solved by using the optimum WMSE solution in equations (\ref{Optimal_equalizer}) and (\ref{Optimal_weights}).

\emph{2) Subproblem over $\big\{\mathbf{W}\big\}$}: 
Since the digital beamformer $\mathbf{W}$ only appears in the second term of the objective function,  (\ref{linear_p5}) is reduced to
\begin{align}\label{linear_p8}
&\min_{\mathbf{W} } ||\mathbf{P}-\mathbf{FW} +\rho\mathbf{D}||^2,
\end{align} 
 which is a quadratic function over $\mathbf{W}$. Therefore, solving $\frac{\partial ||\mathbf{P}-\mathbf{FW}+\rho\mathbf{D}||^2}{\partial \mathbf{W}}=0$ yields the  optimal $\mathbf{W}$ as follows:  
\begin{equation}
\mathbf{W}^o=\left(\mathbf{F}^H\mathbf{F}\right)^{-1}\mathbf{F}^H\left(\mathbf{P}+\rho\mathbf{D}\right).
\label{Optimal_digital}
\end{equation} 

\emph{3) Subproblem over $\big\{\mathbf{F}\big\}$}:  When other blocks are fixed, problem (\ref{linear_p5}) reduces to 
\begin{subequations}\label{linear_p10}
	\begin{align}
&\min_{\mathbf{F} }\text{Tr}\left(\mathbf{F}^H\mathbf{F}\mathbf{Y}\right)-2\text{Re}\left(\text{Tr}\left(\mathbf{F}^H\mathbf{Z}\right)\right),\label{ob_a10}\\
	\text{s.t.}~
 &|\mathbf{F}_{n,i}|=1,~n\in\mathcal{N}_{\text{t}},~i\in\mathcal{N}_{\text{f}},\label{ob_b10}
	\end{align}
\end{subequations}
where $\mathbf{Y}=\mathbf{W}\mathbf{W}^H$ and $\mathbf{Z}=\left(\mathbf{P}+\rho\mathbf{D}\right)\mathbf{W}^H$. It is easily observed that the elements of $\mathbf{F}$ are separated in the unit-modulus constraint (\ref{ob_b10}). This motivates us to optimize $\mathbf{F}$ in an element-wise manner. Specifically, we optimize $\mathbf{F}_{n,i}$ while keeping the remaining elements as their previous values. Consequently, the subproblem of optimizing $\mathbf{F}_{n,i}$ is given by
\begin{subequations}\label{linear_p11}
	\begin{align}
&\min_{\mathbf{F}_{n,i}}\phi_{n,i}|\mathbf{F}_{n,i}|^2-2\text{Re}\left(\chi_{n,i}\mathbf{F}_{n,i}\right),\label{ob_a11}\\
\text{s.t.}~
 &|\mathbf{F}_{n,i}|=1.\label{ob_b11}
\end{align}
\end{subequations}
where $\phi_{n,i}$ and $\chi_{n,i}$ are respectively real and complex constant coefficients, which are determined by the elements of $\mathbf{F}$ except for $\mathbf{F}_{n,i}$. Due to $|\mathbf{F}_{n,i}|=1$, problem (\ref{linear_p11}) is equivalent to $ \max_{\mathbf{F}_{n,i}}2\text{Re}\left(\chi_{n,i}\mathbf{F}_{n,i}\right)$ under the unit-modulus constraint. As a result, the optimal $\mathbf{F}_{n,i}$ can be derived by
\begin{equation}
\mathbf{F}^o_{n,i}=\frac{\chi^*_{n,i}}{|\chi_{n,i}|}.
\label{Optimal_analog}
\end{equation} 
However, the expression of $\chi_{n,i}$ in (\ref{Optimal_analog}) remains unknown. To derive its specific value, we observe that the objective functions of problems (\ref{linear_p10}) and (\ref{linear_p11}) should have equal partial derivatives to $\mathbf{F}_{n,i}$. Consequently, we have 
\begin{equation}
\mathbf{X}_{n,i} -\mathbf{Z}_{n,i} = \phi_{n,i}\mathbf{F}^{(t)}_{n,i}-\chi_{n,i},
\end{equation} 
where $\mathbf{X} = \mathbf{F}^{(t)}\mathbf{Y}$ and $\mathbf{F}^{(t)}$ denotes that the optimized solution of $\mathbf{F}$ in the last iteration. Moreover, we can derive $\phi_{n,i}\mathbf{F}^{(t)}_{n,i}=\mathbf{F}^{(t)}_{n,i}\mathbf{Y}_{i,i}$ by expanding $\mathbf{F}^{(t)}\mathbf{Y}$, so we can obtain
\begin{equation}
\chi_{n,i}=\mathbf{Z}_{n,i}-\mathbf{X}_{n,i} + \mathbf{F}^{(t)}_{n,i}\mathbf{Y}_{i,i}.
\end{equation}

\begin{algorithm}[t]
	\caption{BCD algorithm for solving AL problem (\ref{linear_p5})}
	\begin{algorithmic}[1]\label{Alg.2}
		\STATE Set $\mathbf{F}^{(1)}=\mathbf{F}^{(n-1)}$ and $\mathbf{W}^{(1)}=\mathbf{W}^{(n-1)}$,  initialize $\mathbf{U}^{(1)}$, set iteration index $t=1$,  and the maximum tolerance.
		\WHILE{ not convergent}
		\STATE Update $\omega^{(t)}_{c,k}$ and $\omega^{(t)}_{p,k}$ according to equation (\ref{Optimal_equalizer}).
        \STATE Update $\eta^{(t)}_{c,k}$ and $\eta^{(t)}_{p,k}$ according to equation (\ref{Optimal_weights}).
        \STATE Update $\mathcal{Q}^{(t)}_1$ by solving problem (\ref{linear_p7}).
        \STATE Update $\mathbf{W}^{(t+1)}$ according to equation (\ref{Optimal_digital}).
        \STATE Update $\mathbf{F}^{(t+1)}$ according to equation (\ref{Optimal_analog}).
		\STATE Update iteration index $t=t+1$.
		\ENDWHILE	
		\STATE Output $\mathbf{P}^{(n)}$, $\mathbf{F}^{(n)}$, and $\mathbf{W}^{(n)}$, which are fed back to line 3 of Alg.~\ref{Alg.1}.
	\end{algorithmic}
\end{algorithm}

We summarize the proposed BCD algorithm in Alg.~\ref{Alg.2}. Here are its crucial properties, including convergence and complexity.
\begin{itemize}
\item  \emph{Convergence}: From an arbitrary feasible initial point, our algorithm always seeks the optimal solution in lines $3\sim 8$. We thus deduce that it can find the last feasible point at least after each iteration. Let $f\left(\mathcal{Q}^{(t)}_1,\mathbf{W}^{(t)},\mathbf{F}^{(t)}\right)$ denote the objective function value at $t$-th iteration. we can derive
\begin{align}
&f\left(\mathcal{Q}^{(t-1)}_1,\mathbf{W}^{(t-1)},\mathbf{F}^{(t-1)}\right)\geq f\left(\mathcal{Q}^{(t)}_1,\mathbf{W}^{(t-1)},\mathbf{F}^{(t-1)}\right)\notag\\&\geq f\left(\mathcal{Q}^{(t)}_1,\mathbf{W}^{(t)},\mathbf{F}^{(t-1)}\right)\geq f\left(\mathcal{Q}^{(t)}_1,\mathbf{W}^{(t)},\mathbf{F}^{(t)}\right)
 \label{convergence}
\end{align}
Meanwhile, since objective function value cannot be reduced infinitely, the proposed BCD algorithm converges within a finite number of iterations.

\item \emph{Complexity}: The main complexity stems from solving the problem (\ref{linear_p7}) in line 5 and matrix multiplication operation. The complexity of the interior point method is $\mathcal O\left(N_v^{3.5}\right)$, where $N_v$ is the number of variables. Therefore, the complexity of line 5 is $\mathcal O\left(\left(N_{\text{t}}(N_{\text{t}}+K)+M^2\right)^{3.5}\right)$. For two matrices $\mathbf{W}_1\in\mathbb C^{ A_1\times A_2}$ and $\mathbf{W}_2\in\mathbb C^{ A_2\times A_3}$, the complexity of $\mathbf{W}_1\mathbf{W}_2$ is $\mathcal O\left(A_1A_2A_3\right)$. Therefore, the complexity of lines 3, 4, 6, and 7 are in order of $\mathcal O\left(KN^2_t\right)$, $\mathcal O\left(KN^2_t\right)$, $\mathcal O\left(N_{\text{t}}N_{\text{f}}\max\left(N_{\text{f}}, K+1\right)\right)$, and $\mathcal O\left(N_{\text{t}}N_{\text{f}}(K+1)\right)$, respectively. By retaining the higher-order terms, the per-iteration computational complexity of Alg.~\ref{Alg.2} relies on $\mathcal O\left(\left(N_{\text{t}}(N_{\text{t}}+K)+M^2\right)^{3.5}\right)$.
\end{itemize}

\subsection{Low-complexity HAD beamforming design}
A PDD-based double-loop algorithm was proposed in the previous two subsections to optimize HAD beamformers. However, NF-ISAC typically involves ultra-large antennas, leading to significant computational burdens on the BS when using the double-loop algorithm. To mitigate this issue, a heuristic two-stage HAD beamforming design framework is now introduced, eliminating the need for double-loop iteration. The analog beamformer is designed to maximize the array gain for communication users in this framework. Subsequently, the digital beamformers are optimized based on the designed analog beamformer. Specifically, as the NLoS paths experience double path loss, the LoS path becomes the dominant factor in communication links within NF-ISAC networks. Therefore, maximizing the array gain of the LoS link is considered. To ensure fairness among users, the dimensions of the analog beamforming matrix are evenly distributed. The remaining ${\rm{mod}}\left(N_{\text{f}}, K\right)$ analog beamforming vectors are designed to enhance the array gain of all communication users. Let $\mathbf{f}_j$ be the $j$-th column vector of matrix $\mathbf{F}$. Therefore, the analog beamformer aligning to the array response vector can be given by
\begin{equation}\label{fra}
\mathbf{f}_{j}=
\begin{cases}
\mathbf{a}\left(d_k,\theta_k\right),&\mbox{if $1\leq j\leq \left\lfloor \frac{N_{\text{f}}}{K}\right\rfloor K$ };\\
\frac{1}{|\tilde{\mathbf{a}}|}\odot \tilde{\mathbf{a}}, &\mbox{if $\left\lfloor \frac{N_{\text{f}}}{K}\right\rfloor K+1\leq j\leq N_{\text{f}}$ }.
\end{cases}
\end{equation}
where $k={\rm{mod}}\left(j,K\right)+1$ and $\tilde{\mathbf{a}}=\sum_{k=1}^{K}\mathbf{a}\left(d_k,\theta_k\right)$. With the known analog beamformer, the optimization problem (\ref{linear_p2}) can be simplified as
\begin{subequations}\label{linear_p9}
	\begin{align}
&\min_{\mathbf{W},\mathbf{c} } \text{Tr}\left(\text{CRB}\left(\mathbf{r},\bm{\theta}\right)\right),\label{ob_a9}\\
	\text{s.t.}~
 &\mbox{ (\ref{ob_b}),~(\ref{ob_c}),~(\ref{ob_d}),~(\ref{ob_b2})}.\label{ob_b9} 
	\end{align}
\end{subequations}
The problem (\ref{linear_p9}) appears intractable due to the coupled digital beamformers and fractional SINR.  This motivates us to employ the WMMSE method to recast the communication rate. Based on the WMMSE principle, constraints (\ref{ob_c}) and (\ref{ob_b2}) in problem (\ref{linear_p9}) can be transformed to (\ref{ob_b7}) and (\ref{ob_c7}). However, $T_{c,k}$, $I_{c,k}$, $T_{p,k}$, and $I_{p,k}$ should be updated to 
\begin{equation}
T_{c,k}=\overbrace{{\left|\tilde{\mathbf{h}}_{k}\mathbf{w}_0\right|}^2}^{S_{c,k}}+\underbrace{\overbrace{{\left|\tilde{\mathbf{h}}_{k}\mathbf{w}_{k}\right|}^2}^{S_{{p,k}}}+\overbrace{\sum_{j=1,j\neq k}^{K}\left|\tilde{\mathbf{h}}_{k}\mathbf{w}_j\right|^2+\sigma^2}^{I_{{p,k}}}}_{I_{c,k}=T_{p,k}},
\label{equ:rece_power3}
\end{equation}
where $\tilde{\mathbf{h}}_{k} =\mathbf{h}^H_k\mathbf{F}$. Then, to recast the FIM, we introduce an auxiliary matrix $\mathbf{Q}=\mathbf{F}\mathbf{W}\mathbf{W}^H\mathbf{F}^H$. Using the Schur complement, the quadratic equality can be divided into 
\begin{subequations}\label{equation_constraint_2}
\begin{align}
&\begin{bmatrix} \mathbf{Q} & \mathbf{FW} \\ (\mathbf{FW})^H & \mathbf{I} \end{bmatrix}\succeq\mathbf{0},\label{equation_3}\\
 &\text{Tr}\left(\mathbf{Q}-\mathbf{FW}(\mathbf{FW})^H\right)\leq 0,\label{equation_4}
\end{align}
\end{subequations}
Based on the first-order Taylor expansion, the inequality (\ref{equation_4}) can be recast as
\begin{align}
2\text{Re}\left(\text{Tr}\left(\mathbf{F}\mathbf{W}^{(t)}(\mathbf{FW})^H\right)\right)-&\text{Tr}\left(\mathbf{FW}^{(t)}\left(\mathbf{FW}^{(t)}\right)^H\right)\notag\\&\geq \text{Tr}\left(\mathbf{Q}\right),
\label{Surrogate_2}
\end{align}
After introducing auxiliary matrix $\mathbf{U}$ for $\text{Tr}\left(\text{CRB}\left(\mathbf{r},\bm{\theta}\right)\right)$,  the problem (\ref{linear_p11}) can be transformed to
\begin{subequations}\label{linear_p12}
	\begin{align}
&\min_{\tilde{\mathcal{Q}}_1,\tilde{\mathcal{Q}}_2} \text{Tr}\left(\mathbf{U}^{-1}\right),\label{ob_a12}\\
	\text{s.t.}~
  &\mbox{ (\ref{ob_b}),~(\ref{ob_d}),~(\ref{Schur_complement}),~(\ref{ob_b7}),~(\ref{ob_c7}),~(\ref{equation_3}),~(\ref{Surrogate_2})}.\label{ob_b12}
	\end{align}
\end{subequations}
where $\tilde{\mathcal{Q}}_1=\big\{\mathbf{U},\mathbf{Q},\mathbf{W},\mathbf{c}\big\}$ and $\mathcal {Q}_2=\big\{\eta_{c,k}, \omega_{c,k}, \eta_{p,k}, \omega_{p,k}\big\}$. Similar to the solution for the first subproblem in Section \ref{Section III}.B, $\tilde{\mathcal {Q}}_1$ and $\tilde{\mathcal {Q}}_2$ are iteratively updated. The detailed process is summarized in Alg.~\ref{Alg.3}. The per-iteration complexity has the same level as Alg.~\ref{Alg.2}, but it avoids double-loop iteration, which thus is more efficient than the proposed PDD-based double-loop algorithm in practice.
\begin{algorithm}[t]
	\caption{Two-stage HAD beamforming design for solving problem (\ref{linear_p2})}
	\begin{algorithmic}[1]\label{Alg.3}
		\STATE Initialize $\mathbf{W}^{(0)}$, set iteration index $t=1$  and the maximum tolerance.
		\WHILE{ not convergent}
		\STATE Update $\omega^{(t)}_{c,k}$ and $\omega^{(t)}_{p,k}$ according to equation (\ref{Optimal_equalizer}).
        \STATE Update $\eta^{(t)}_{c,k}$ and $\eta^{(t)}_{p,k}$ according to equation (\ref{Optimal_weights}).
        \STATE Update $\tilde{\mathcal{Q}}^{(t)}_1$ by solving problem (\ref{linear_p12}).
		\STATE Update iteration index $t=t+1$.
		\ENDWHILE	
		\STATE Output the optimized sensing performance.
	\end{algorithmic}
\end{algorithm} 

\section{Extension to partially-connected phase shift architecture}\label{Section IV}
This section focuses on the HAD beamforming design for the partially connected phase shift architecture, which can be solved using the proposed PDD-based double-loop algorithm.

We first define  a diagonal matrix $\tilde{\mathbf{F}}\in\mathbb{C}^{N_{\text{t}}\times N_{\text{t}}}$, as follows, 
\begin{equation}
\tilde{\mathbf{F}}=\text{Bdiag}\left(\text{diag}\left(\mathbf{f}_1\right),\dots,\text{diag}\left(\mathbf{f}_{N_{\text{f}}}\right)\right),
\label{diagonal}
\end{equation}
where $\text{Bdiag}\left( \bullet\right)$ operator is defined in \emph{Notations} and  each diagonal element of $\tilde{\mathbf{F}}$ meets the unit-modulus constraint. The block-diagonal matrix $\mathbf{F}$ can then  be expressed by 
\begin{equation}
\mathbf{F} = \tilde{\mathbf{F}}\mathbf{\Phi} = \text{diag}\left(\mathbf{f}\right)\mathbf{\Phi},
\label{A}
\end{equation}
where $\mathbf{f}=\left[\mathbf{f}^T_1,\dots,\mathbf{f}^T_{N_{\text{f}}}\right]^T$ and transformation matrix $\mathbf{\Phi}$ is given by
\begin{equation}
\mathbf{\Phi} =\left[{\boldsymbol{\phi}}_1,\dots,\boldsymbol{\phi}_{N_{\text{f}}}\right]=\text{Bdiag}\left(\mathbf{1},\dots,\mathbf{1}\right),
\label{A}
\end{equation} where $\boldsymbol{\phi}_i$ denotes the $i$-th column of the transformation matrix for $\forall i\in\mathcal{N}_{\text{f}}$ and $\mathbf{1}$ is a $N_{\text{c}}\times 1$ vector with all elements being 1. Then, following the same procedure as the PDD framework in Section \ref{Section III}.A, the AL problem can be formulated as
\begin{subequations}\label{linear_p13}
	\begin{align}
&\min_{\mathbf{P},\mathbf{Q},\tilde{\mathbf{F}},\mathbf{W},\mathbf{U},\mathbf{c} } \text{Tr}\left(\mathbf{U}^{-1}\right)+\frac{1}{2\rho}||\mathbf{P}-\tilde{\mathbf{F}} \mathbf{\Phi W}+\rho\mathbf{D}||^2,\label{ob_a13}\\
	\text{s.t.}~
 &|\tilde{\mathbf{F}}_{n,n}|=1,~\forall n\in\mathcal{N}_{\text{t}},\label{ob_b13}\\
  &\mbox{(\ref{ob_c}),~(\ref{ob_d})
  ,~(\ref{ob_b2}),~(\ref{equation_constraint}),~(\ref{Schur_complement}),~(\ref{ob_c4})}.\label{ob_c13}
	\end{align}
\end{subequations}
We divide the optimization variables into three blocks and then adopt the BCD method to optimize these blocks alternately\cite{10587118}.

\emph{1) Subproblem over $\left\{\mathbf{P},\mathbf{Q},\mathbf{U},\mathbf{c}\right\}$}: Since this does not involve $\tilde{\mathbf{F}}$, it can be solved same as the first subproblem in Section \ref{Section III}.B. The details are omitted for brevity.

\emph{2) Subproblem over $\left\{\mathbf{W}\right\}$}: When other variable blocks are fixed, optimizing $\mathbf{W}$ can be simplified as an unconstrained problem, which is given by
\begin{align}\label{linear_p14}
&\min_{\mathbf{W} } ||\mathbf{P}-\tilde{\mathbf{F}} \mathbf{\Phi W} +\rho\mathbf{D}||^2.
\end{align}
By setting the first-order partial derivative of the above objective over $\mathbf{W}$ to zero, the optimal digital beamformer can be derived as 
\begin{equation}
\mathbf{W}^o=\left(\left(\tilde{\mathbf{F}} \mathbf{\Phi }\right)^H\tilde{\mathbf{F}} \mathbf{\Phi }\right)^{-1}\left(\tilde{\mathbf{F}} \mathbf{\Phi }\right)^H\left(\mathbf{P}+\rho\mathbf{D}\right).
\label{Optimal_digital_2}
\end{equation} 

\emph{3) Subproblem over $\big\{\tilde{\mathbf{F}}\big\}$}: The analog beamformer $\tilde{\mathbf{F}}$ with unit-modulus diagonal element constraint only appears in the last term of the objective function, so the subproblem can be formulated as
\begin{subequations}\label{linear_p15}
	\begin{align}
&\min_{\tilde{\mathbf{F}} }\text{Tr}\left(\tilde{\mathbf{F}} ^H\tilde{\mathbf{F}} \tilde{\mathbf{Y}}\right)-2\text{Re}\left(\text{Tr}\left(\tilde{\mathbf{F}} ^H\tilde{\mathbf{Z}}\right)\right),\label{ob_a15}\\
	\text{s.t.}~
 &|\tilde{\mathbf{F}}_{n,n}|=1,~n\in\mathcal{N}_{\text{t}},\label{ob_b15}
	\end{align}
\end{subequations}
where $\tilde{\mathbf{Y}}=\mathbf{\Phi}\mathbf{W}\mathbf{W}^H\mathbf{\Phi}^H$ and $\tilde{\mathbf{Z}}=\left(\mathbf{P}+\rho\mathbf{D}\right)\mathbf{W}^H\mathbf{\Phi}^H$. We next design $\tilde{\mathbf{F}}$ in an element-wise manner, which optimizes each entry of $\tilde{\mathbf{F}}$ while fixing the remaining elements.  As a result, optimizing $\mathbf{F}_{n,n}$ can be given by 
\begin{subequations}\label{linear_p16}
	\begin{align}
&\min_{\mathbf{F}_{n,n}}\tilde{\phi}_{n,n}|\tilde{\mathbf{F}}_{n,n}|^2-2\text{Re}\left(\tilde{\chi}_{n,n}\tilde{\mathbf{F}}_{n,n}\right),\label{ob_a16}\\
\text{s.t.}~
 &|\tilde{\mathbf{F}}_{n,n}|=1.\label{ob_b16}
\end{align}
\end{subequations}
Following the similar derivation of solving problem (\ref{linear_p11}), we can get the optimal $\tilde{\mathbf{F}}_{n,n}$, which is given by
\begin{equation}
\tilde{\mathbf{F}}^o_{n,n}=\frac{\tilde{\chi}^*_{n,n}}{|\tilde{\chi}_{n,n}|},
\label{Optimal_analog_2}
\end{equation} 
with 
\begin{equation}
\tilde{\chi}_{n,n}=\tilde{\mathbf{Z}}_{n,n}-\tilde{\mathbf{X}}_{n,n} + \tilde{\mathbf{F}}^{(t)}_{n,n}\tilde{\mathbf{Y}}_{n,n}.
\end{equation}
where $\tilde{\mathbf{X}} = \tilde{\mathbf{F}}^{(t)}\tilde{\mathbf{Y}}$ and $\tilde{\mathbf{F}}^{(t)}$ denotes that the optimized solution of $\tilde{\mathbf{F}}$ in the last iteration. The proposed BCD algorithm for partially-connected phase shift architecture is summarized in Alg.~\ref{Alg.4}. Its complexity is same as the Alg.~\ref{Alg.2}.

\begin{algorithm}[t]
	\caption{BCD algorithm for solving AL problem (\ref{linear_p13})}
	\begin{algorithmic}[1]\label{Alg.4}
		\STATE Set $\mathbf{F}^{(1)}=\mathbf{F}^{(n-1)}$ and $\mathbf{W}^{(1)}=\mathbf{W}^{(n-1)}$,  initialize $\mathbf{U}^{(1)}$, set iteration index $t=1$,  and the maximum tolerance.
		\WHILE{ not convergent}
		\STATE Call lines 3$\sim$5 of Alg.~\ref{Alg.2}.
        \STATE Update $\mathbf{W}^{(t+1)}$ according to equation (\ref{Optimal_digital_2}).
        \STATE Update $\mathbf{F}^{(t+1)}$ according to equation (\ref{Optimal_analog_2}).
		\STATE Update iteration index $t=t+1$.
		\ENDWHILE	
		\STATE Output $\mathbf{P}^{(n)}$, $\mathbf{F}^{(n)}$, and $\mathbf{W}^{(n)}$, which are fed back to line 3 of Alg.~\ref{Alg.1}.
	\end{algorithmic}
\end{algorithm}

\begin{table}[t]
	\caption{Key simulation parameters}
        \vspace{-0.2cm}
	\begin{center}\label{Table II}
		\begin{tabular}{|l||l| p{13cm}}       
            \hline
			\bf{Parameter} & \bf{Value} \\
			\hline
			Number of transmit antennas& $N_{\text{t}}=64$\\
               \hline
			Number of receive antennas& $N_{\text{r}}=32$\\
            \hline
			Number of RF chains& $N_{\text{f}}=8$\\
            \hline
			Antenna array aperture& $N_{\text{t}}=N_{\text{r}}=0.5$~m\\
            \hline
			Rayleigh distance& $50$~m\\
            \hline
			Number of communication users& $K=4$\\
            \hline
			Minimum communication requirement & $R_{\text{th}}=2$~bps/Hz\\
            \hline
			Number of targets& $M=2$\\
            \hline
			Number of NLoS paths& $Q=2$\\
                \hline
			Carrier frequency & $f_{\text{c}}=30$~GHz\\
			\hline
                Maximum transmit power &$P_{\text{th}}=30$~dBm\\
                \hline
                Background noise &$\sigma^2=-90$~dBm\\
			\hline
		\end{tabular}
	\end{center}
\end{table}

\section{Simulation result}\label{Section V}

Comprehensive simulations evaluate the proposed transmit scheme and algorithms in a system with randomly distributed users and targets. Targets are positioned within $20\sim 30$~m, while users and targets are spread across the entire NF region. Results are averaged over 100 independent NF channel realizations. Unless stated otherwise, Table \ref{Table II} details simulation parameters, primarily based on \cite{10579914,10520715}.

Our proposed Fully-Connected, Low-Complexity, and Partially-Connected HAD optimization algorithms for RSMA-enabled NF-ISAC are respectively labeled as {\bf{RSMA-FC-near}}, {\bf{RSMA-LC-near}}, and  {\bf{RSMA-PC-near}}. To comprehensively assess their performance, we benchmark them against three baselines, as follows,
\begin{itemize}
\item {\bf{RSMA-FD-near}}: The BS adopts the fully digital beamforming structure, where each antenna has a dedicated RF chain. This benchmark provides the CRB lower bound for our proposed fully-connected and partially-connected HAD beamforming structures.
\item {\bf{SDMA-FC-near}}: This scheme uses the fully-connected HAD beamforming. However, the message intended for each user is encoded into one stream while each communication user decodes its desired stream by treating other streams as additional noise.
\item {\bf{RSMA-FC-far}}:  Communication channels adopt a far-field plane-wave propagation model, where the far-field array response vector is in equation (\ref{Response_vector}) is updated to
\begin{equation}
\mathbf{a}_{\text{far}}\left(\theta\right)= \left[e^{j\frac{2\pi}{\lambda}d\sin\theta},\dots,e^{j\frac{2\pi}{\lambda}Nd\sin\theta}\right]^T.
\label{Far-Channel}
\end{equation}
Except for the array response vector, other parameters remain identical to ensure comparison fairness. However, sensing channels still adopt the NF spherical-wave propagation model since far-field channels cannot simultaneously resolve angle and distance. This setting overstates the performance of far-field ISAC.
\end{itemize}

\begin{figure}[tbp]
	\centering
	\subfigure[Sum RCRB of angle versus the minimum rate requirement]{
		\begin{minipage}[t]{0.46\linewidth}
			\centering\includegraphics[width = 1.58in]{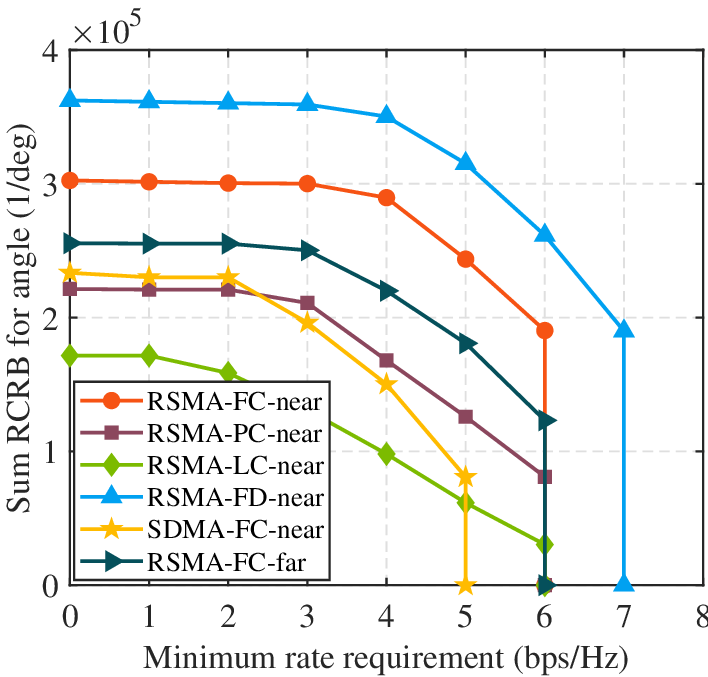}
		\end{minipage}
	}
	\subfigure[Sum RCRB of distance versus the minimum rate requirement]{
		\begin{minipage}[t]{0.47\linewidth}
			\centering\includegraphics[width = 1.58in]{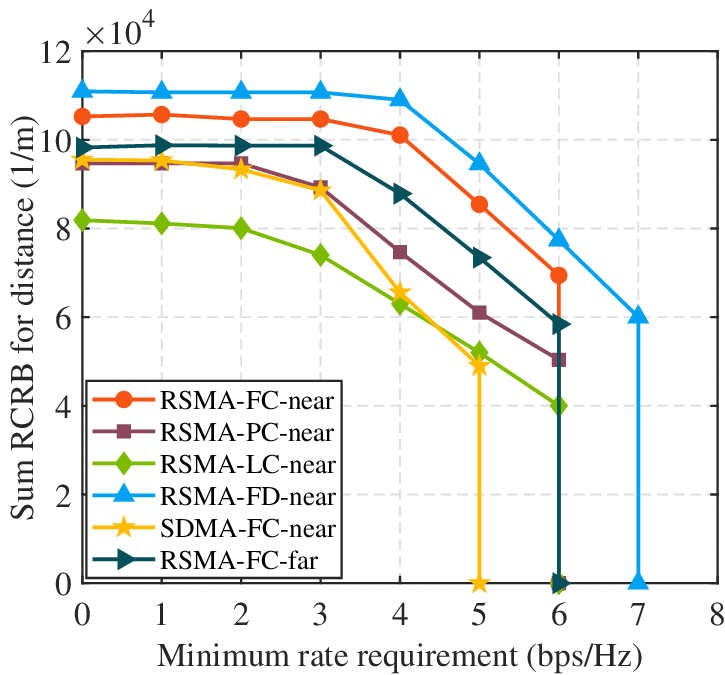}
		\end{minipage}
	}
	\centering
	\caption{RCRB versus the minimum rate requirement}
	\label{Fig_QoS}
    \vspace{-0.5cm}
\end{figure}
Fig.~\ref{Fig_QoS} illustrates the Pareto boundaries of the sum RCRB versus the minimum communication rate requirement. The proposed fully connected HAD beamformer with eight RF chains achieves sensing performance comparable to a fully digital beamformer with $64$ RF chains, particularly in multi-target distance sensing. Compared to FF ISAC and SDMA-aided NF-ISAC, the RSMA-enabled NF-ISAC achieves superior tradeoff performance and a larger achievable performance region.
This improvement is attributed to two key factors. First, the NF beamformer concentrates energy on specific points, effectively mitigating intra-user interference. Second, RSMA introduces an additional beamforming vector for the common stream, providing $K+1$ digital beamformers for a 
$K$-user ISAC network, offering greater flexibility for sensing. The common stream also manages interference among communication users and between sensing and communication functions.
Notably, SDMA's sensing performance degrades rapidly as communication rate requirements increase. And it supports a maximum rate of only 5~bps/Hz, whereas RSMA achieves 6~bps/Hz, highlighting RSMA’s superior adaptability in managing interference.

\begin{figure}[tbp]
	\centering
	\subfigure[Sum RCRB of angle versus transmit power threshold]{
		\begin{minipage}[t]{0.46\linewidth}
			\centering\includegraphics[width = 1.58in]{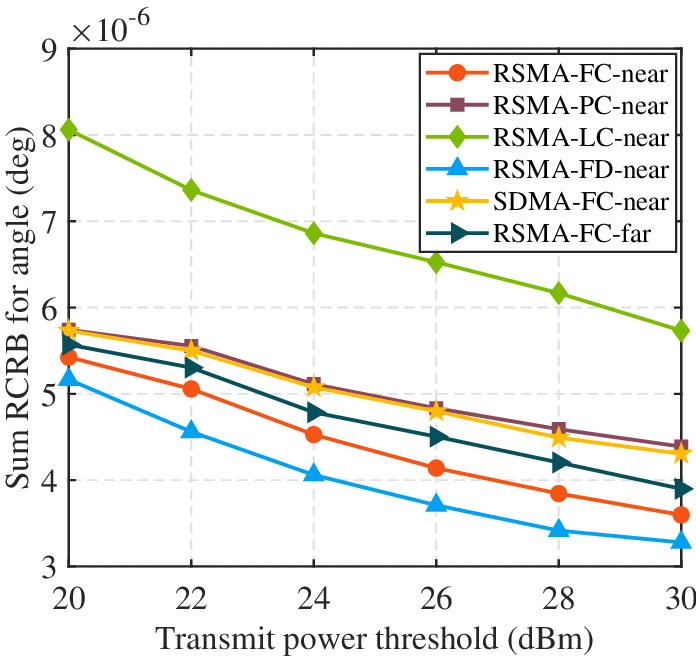}
		\end{minipage}
	}
	\subfigure[Sum RCRB of distance versus transmit power threshold]{
		\begin{minipage}[t]{0.46\linewidth}
			\centering\includegraphics[width = 1.58in]{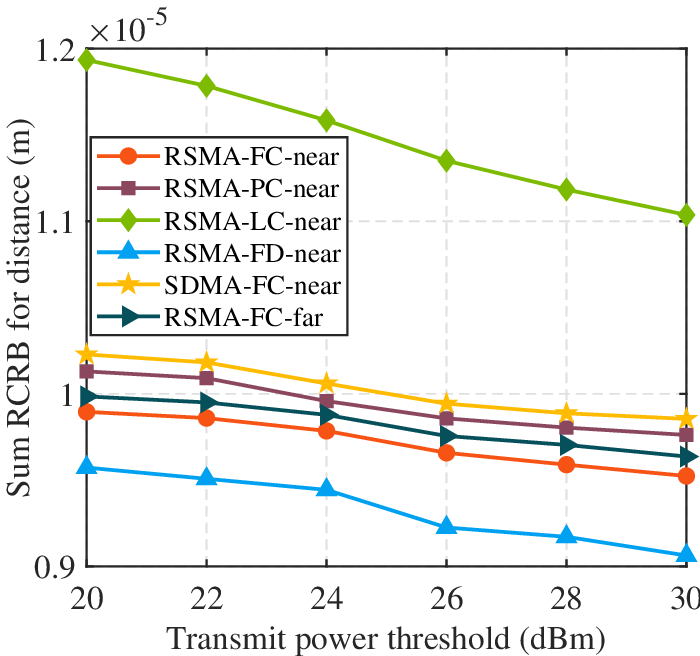}
		\end{minipage}
	}
	\centering
	\caption{RCRB versus transmit power threshold}
	\label{Fig_power}
    \vspace{-0.3cm}
\end{figure}

Fig.~\ref{Fig_power} illustrates the sum-RCRB for angle and distance sensing versus the transmit power threshold. Three observations can be made. First, as the transmit power increases, the sensing error of all transmit approaches and algorithms decreases since the increased transmit power broadens the solution space. Second, our proposed transmit scheme is consistently superior to FF ISAC and SDMA-assisted NF-ISAC, achieving $10\%$ and $17\%$ angle sensing gains through more effective interference management. Third, compared to the fully digital beamformer, the fully connected and partially connected HAD beamformers lose approximately $10\%$ and $25\%$ of angle sensing performance. However, the number of RFs has decreased by eight times. This suggests the effectiveness of the proposed algorithm. 

\begin{figure}[tbp]
	\centering
	\subfigure[Sum RCRB of angle versus the number of RF chains]{
		\begin{minipage}[t]{0.46\linewidth}
			\centering\includegraphics[width = 1.58in]{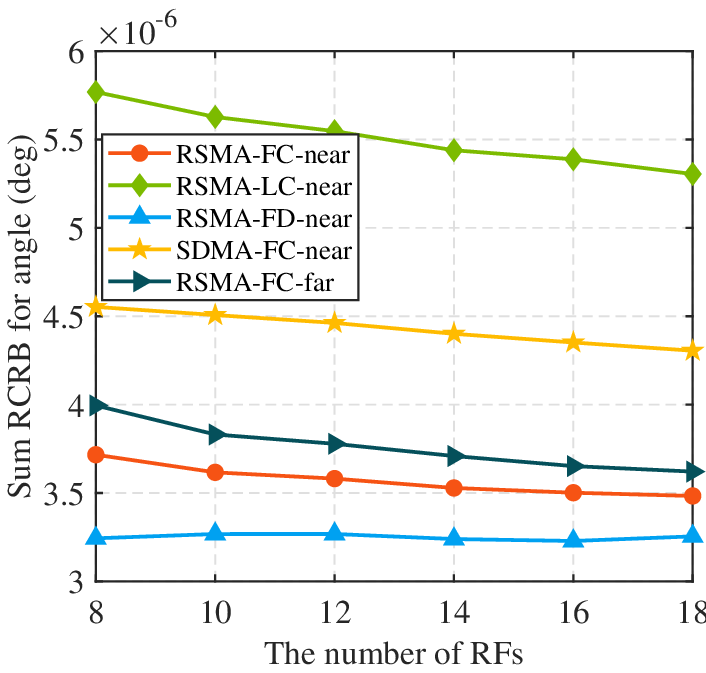}
		\end{minipage}
	}
	\subfigure[Sum RCRB of distance versus the number of RF chains]{
		\begin{minipage}[t]{0.46\linewidth}
			\centering\includegraphics[width = 1.58in]{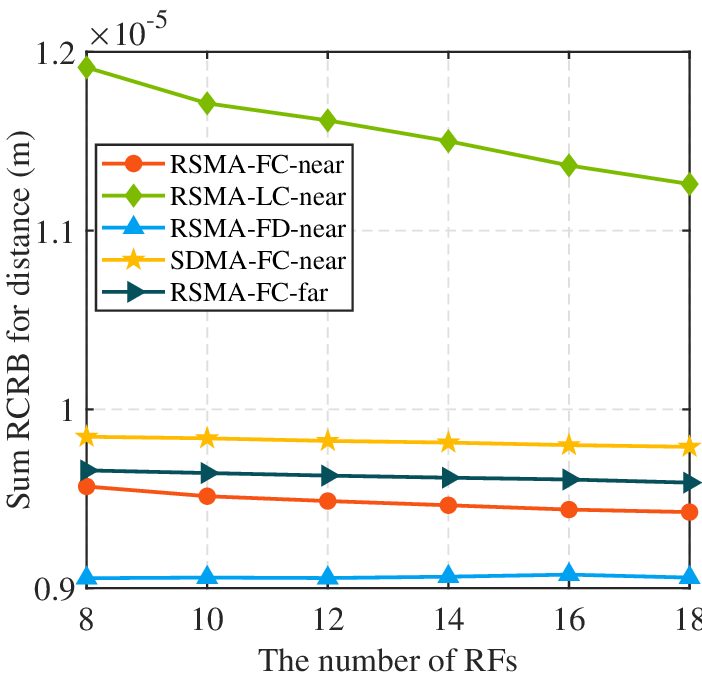}
		\end{minipage}
	}
	\centering
	\caption{RCRB versus the number of RF chains}
	\label{Fig_RF}
    \vspace{-0.3cm}
\end{figure}
Fig.~\ref{Fig_RF} shows the sum-RCRB for angle and distance sensing versus the number of RF chains. The sensing performance of the fully-digital beamforming structure remains constant, serving as a lower bound for fully-connected and partially-connected HAD beamformers. As the number of RF chains increases, the performance gap between fully-digital and HAD beamforming structures narrows. For instance, at $N_{\text{f}}=18$, the gaps for angle and distance sensing reduce to $0.2\times 10^{-6}$ and $0.5\times 10^{-6}$, respectively. This improvement arises because the digital beamforming dimension becomes sufficient to mitigate intra-user interference as RF chains increase. Notably, the proposed RSMA-based transmit scheme maintains its performance gains compared to FF ISAC and SDMA.

\begin{figure}[tbp]
	\centering
	\subfigure[Sum RCRB of angle versus the number of users]{
		\begin{minipage}[t]{0.46\linewidth}
			\centering\includegraphics[width = 1.58in]{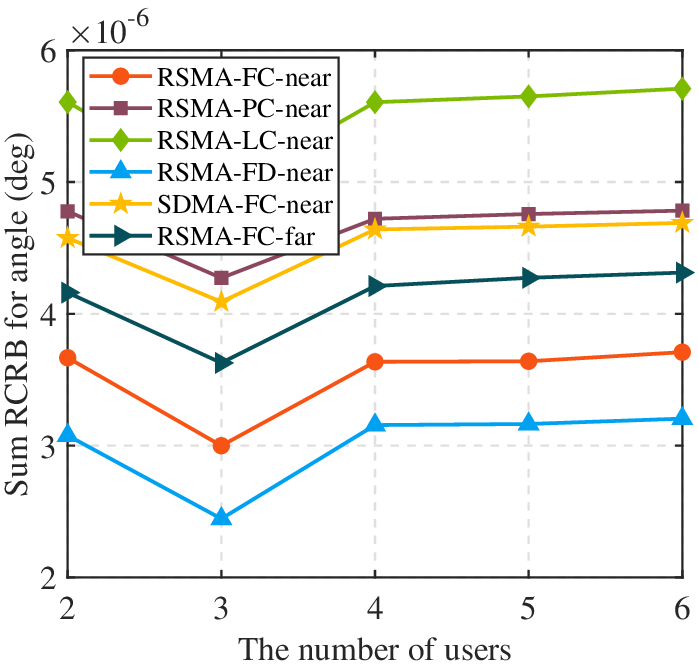}
		\end{minipage}
	}
	\subfigure[Sum RCRB of distance versus the number of users]{
		\begin{minipage}[t]{0.46\linewidth}
			\centering\includegraphics[width = 1.58in]{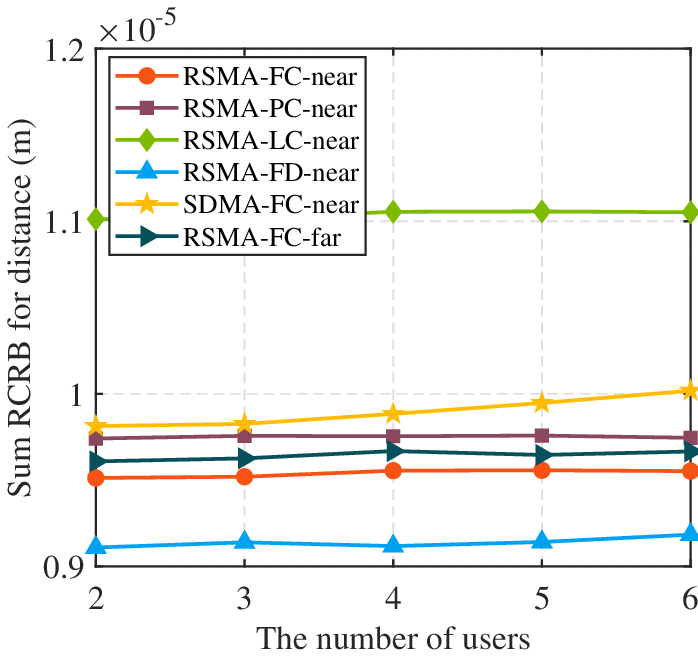}
		\end{minipage}
	}
	\centering
	\caption{RCRB versus the number of users}
	\label{Fig_user}
    \vspace{-0.3cm}
\end{figure}
Fig.~\ref{Fig_user} simulates the sum-RCRB for angle and distance sensing versus the number of users. The result shows that the sensing error increases slightly as a whole as the number of users increases. An interesting observation is that the angle sensing error of all transmission schemes and algorithms reduces when $K=3$. This may be because the communication performance is still easily satisfied due to the weak intra-user interference. In this case, the number of beamforming vectors is beneficial for target sensing. However, as the number of users increases, so does the intra-user interference. When $K>3$, the positive effects of digital beamforming quantity have gradually been counteracted by intra-user interference. From Fig.~\ref{Fig_user} (b), the distance sensing performance gap between RSMA and SDMA becomes more noticeable as the number of users increases. This is because the excess interference degrades SDMA's communication performance. Meanwhile, this observation highlights the capability of RSMA in interference management.

\section{Conclusion}\label{Section VI}
An RSMA-based transmit scheme with fully and partially connected HAD beamforming architectures is proposed and investigated for an NF-ISAC network with multiple targets and users. To reveal the fundamental CRB-rate performance tradeoff, we derive CRB for joint multi-target distance and angle estimation and formulate a dual-objective optimization problem. Then, the formulated problem is recast to a sensing-centric non-convex minimization problem with minimum communication rate constraint. By leveraging WMMSE and first-order Taylor expansion, we develop a PDD-based double-loop algorithm to jointly optimize fully-connected HAD beamformer and common rate allocation. To reduce the computational complexity, a two-stage beamforming design algorithm is proposed. The PDD-based double-loop algorithm is extended to partially-connected HAD beamforming scenarios. Simulation results show that: 1) the proposed fully-connected and partially-connected HAD beamformers perform comparably to the fully digital beamformer and 2)  the proposed transmit scheme achieves significant performance gain and enhanced CRB-rate region over far-field ISAC and SDMA.

\appendices
\section{Derivation of CRB matrix}
To derive the CRB with the closed-form expressions, equation (\ref{Sensing_matrix}) is vectorized as 
\begin{equation}
\tilde{\mathbf{y}}=\left(\mathbf{X}^T\otimes\mathbf{I}_{N_{\text{r}}}\right)\tilde{\mathbf{g}}+\tilde{\mathbf{n}},
\label{Vectorized_Sensing}
\end{equation}
where $\tilde{\mathbf{y}}=\text{vec}\left(\mathbf{Y}\right)\in\mathbb{C}^{N_{\text{r}}L\times 1}$, $\tilde{\mathbf{g}}=\text{vec}\left(\mathbf{G}\right)\in\mathbb{C}^{N_{\text{r}}N_{\text{t}}\times 1}$, and $\tilde{\mathbf{n}}=\text{vec}\left(\mathbf{N}_0\right)\in\mathbb{C}^{N_{\text{r}}L\times 1}$.
It can be easily validated that the received signal vector $\tilde{\mathbf{y}}$ is a CSCG random vector with mean $\mathbf{u}$ and covariance $\mathbf{v}$, \emph{i.e.}, $\tilde{\mathbf{y}}\sim \mathcal{CN}\left(\mathbf{u},\mathbf{v}\right)$, where $\mathbf{u}=\left(\mathbf{X}^T\otimes\mathbf{I}_{N_{\text{r}}}\right)\tilde{\mathbf{g}}$ and $\mathbf{v}=\sigma^2_0\mathbf{I}_{N_{\text{r}}L}$. Let $\mathbf{J}$ denote the FIM for estimation $\boldsymbol{\xi}$. Following the similar derivation procedures in \cite{4359542}, the element at the $i$-th row and the $j$-th column of $\mathbf{J}_{\bm{\xi}}$ can be given by 
\begin{align}
\left[\mathbf{J}_{\bm{\xi}}\right]_{i,j}=& 2\text{Re}\left\{\frac{\partial \mathbf{u}^H}{\partial \bm{\xi}_i}\mathbf{v}^{-1}\frac{\partial \mathbf{u}}{\partial \bm{\xi}_j}\right\} + \text{Tr}\left(\mathbf{v}^{-1}\frac{\partial \mathbf{v}^H}{\partial \bm{\xi}_i}\mathbf{v}^{-1}\frac{\partial \mathbf{v}}{\partial \bm{\xi}_j}\right) \notag\\=& \frac{2}{\sigma^2_0}\text{Re}\left\{\frac{\partial \mathbf{u}^H}{\partial \bm{\xi}_i}\frac{\partial \mathbf{u}}{\partial \bm{\xi}_j}\right\},
\end{align}
where $\bm{\xi}_i$ denotes the $i$-th element of $\bm{\xi}$. Consequently, we have
\begin{align}
\mathbf{J}_{\bm{\xi}}&=\frac{2L}{\sigma^2_0}\left[\begin{array}{cc|cc}
\mathbf{J}_{\bm{\theta}\bm{\theta}} & \mathbf{J}_{\bm{\theta}\mathbf{r}} & \mathbf{J}_{\bm{\theta}\bm{\beta}_\text{R}} & \mathbf{J}_{\bm{\theta}\bm{\beta}_\text{I}}\\
\mathbf{J}_{\bm{\theta}\mathbf{r}}&\mathbf{J}_{\mathbf{r}\mathbf{r}}&\mathbf{J}_{\bm{r}\bm{\beta}_\text{R}}&\mathbf{J}_{\bm{r}\bm{\beta}_\text{I}}\\
\hline 
\mathbf{J}_{\bm{\theta}\bm{\beta}_\text{R}}&\mathbf{J}_{\bm{r}\bm{\beta}_\text{R}}&\mathbf{J}_{\bm{\beta}_\text{R}\bm{\beta}_\text{R}}&\mathbf{J}_{\bm{\beta}_\text{R}\bm{\beta}_\text{I}}\\
\mathbf{J}_{\bm{\theta}\bm{\beta}_\text{I}}&\mathbf{J}_{\bm{r}\bm{\beta}_\text{I}}&\mathbf{J}_{\bm{\beta}_\text{R}\bm{\beta}_\text{I}}&\mathbf{J}_{\bm{\beta}_\text{I}\bm{\beta}_\text{I}}
\end{array}\right]\notag\\&=\frac{2L}{\sigma^2_0}\left[\begin{array}{c|c}
\mathbf{J}_{11} & \mathbf{J}_{12}\\
\hline \mathbf{J}^T_{12}&\mathbf{J}_{22}\\
\end{array}\right]
\end{align}  
with  the element in $\mathbf{J}_{\bm{\xi}}$ being calculated as 
\begin{subequations}\label{partial_derivative}
\begin{align}
\mathbf{J}_{\bm{x}\bm{y}}=&\text{Re}\bigg(\left(\dot{\mathbf{A}}^H_{\text{r},\bm{x}}\dot{\mathbf{A}}_{\text{r},\bm{y}}\right)\odot\left(\mathbf{B}^*\mathbf{A}^H_\text{t}\mathbf{R}^*\mathbf{A}_\text{t}\mathbf{B}\right)\notag\\
&+\left(\dot{\mathbf{A}}^H_{\text{r},\bm{x}}\mathbf{A}_{\text{r}}\right)\odot\left(\mathbf{B}^*\mathbf{A}^H_\text{t}\mathbf{R}^*\dot{\mathbf{A}}_{\text{t},\bm{y}}\mathbf{B}\right)\notag\\
&+\left(\mathbf{A}^H_{\text{r}}\dot{\mathbf{A}}_{\text{r},\bm{y}}\right)\odot\left(\mathbf{B}^*\dot{\mathbf{A}}^H_{\text{t},\bm{x}}\mathbf{R}^*\mathbf{A}_\text{t}\mathbf{B}\right)\notag\\
&+\left(\mathbf{A}^H_{\text{r}}\mathbf{A}_{\text{r}}\right)\odot\left(\mathbf{B}^*\dot{\mathbf{A}}^H_{\text{t},\bm{x}}\mathbf{R}^*\dot{\mathbf{A}}_{\text{t},\bm{y}}\mathbf{B}\right)\bigg)\\
\mathbf{J}_{\bm{x}\bm{\beta}_\text{R}}=&\text{Re}\bigg(\left(\dot{\mathbf{A}}^H_{\text{r},\bm{x}}\mathbf{A}_{\text{r}}\right)\odot\left(\mathbf{B}^*\mathbf{A}^H_\text{t}\mathbf{R}^*\mathbf{A}_{\text{t}}\right)\notag\\
&+\left(\mathbf{A}^H_{\text{r}}\mathbf{A}_{\text{r}}\right)\odot\left(\mathbf{B}^*\dot{\mathbf{A}}^H_{\text{t},\bm{x}}\mathbf{R}^*\mathbf{A}_{\text{t}}\right)\bigg)\\
\mathbf{J}_{\bm{x}\bm{\beta}_\text{I}}=&-\text{Im}\bigg(\left(\dot{\mathbf{A}}^H_{\text{r},\bm{x}}\mathbf{A}_{\text{r}}\right)\odot\left(\mathbf{B}^*\mathbf{A}^H_\text{t}\mathbf{R}^*\mathbf{A}_{\text{t}}\right)\notag\\
&+\left(\mathbf{A}^H_{\text{r}}\mathbf{A}_{\text{r}}\right)\odot\left(\mathbf{B}^*\dot{\mathbf{A}}^H_{\text{t},\bm{x}}\mathbf{R}^*\mathbf{A}_{\text{t}}\right)\bigg)\\
\mathbf{J}_{\bm{\beta}_\text{R}\bm{\beta}_\text{R}}=&\mathbf{J}_{\bm{\beta}_\text{I}\bm{\beta}_\text{I}} = \text{Re}\bigg(\left(\mathbf{A}^H_{\text{r}}\mathbf{A}_{\text{r}}\right)\odot\left(\mathbf{A}^H_{\text{t}}\mathbf{R}^*\mathbf{A}_{\text{t}}\right)\bigg)\\
\mathbf{J}_{\bm{\beta}_\text{R}\bm{\beta}_\text{I}}=&-\text{Im}\bigg(\left(\mathbf{A}^H_{\text{r}}\mathbf{A}_{\text{r}}\right)\odot\left(\mathbf{A}^H_{\text{t}}\mathbf{R}^*\mathbf{A}_{\text{t}}\right)\bigg)
\end{align}
\end{subequations}
for any $ \mathbf{x},\mathbf{y}\in\{\bm{\theta},\bm{r}\}$, where
\begin{subequations}
\begin{align}
\dot{\mathbf{A}}_{i,\bm{\theta}} = \left[\frac{\partial \mathbf{a}_i\left(\tilde r_1,\tilde\theta_1\right)}{\partial \tilde\theta_1},\dots,\frac{\partial \mathbf{a}_i\left(\tilde r_M,\tilde\theta_M\right)}{\partial \tilde\theta_M}\right]\\
\dot{\mathbf{A}}_{i,\bm{r}} = \left[\frac{\partial \mathbf{a}_i\left(\tilde r_1,\tilde \theta_1\right)}{\partial \tilde r_1},\dots,\frac{\partial \mathbf{a}_i\left(\tilde r_M,\tilde\theta_M\right)}{\partial \tilde r_M}\right]
\end{align}
\end{subequations}
for any $i\in\{\text{r},\text{t}\}$.

	\ifCLASSOPTIONcaptionsoff
	\newpage
	\fi
	
	\bibliographystyle{IEEEtran}
	\bibliography{references}
	
\end{document}